\documentclass[aps,prb,twocolumn,showpacs,amsmath,amssymb,superscriptaddress]{revtex4}
\usepackage{graphicx}
\usepackage{dcolumn}
\usepackage{amsmath}
\usepackage{amssymb}
\usepackage{color}
\usepackage{epstopdf}
\begin{document}
\title{Quantum phase transitions of a two-leg bosonic ladder in an artificial gauge field}
\author{R. Citro}
\affiliation{Dipartimento di Fisica "E.R. Caianiello", Universit\`a
  degli Studi di Salerno and Unit\`a Spin-CNR, Via Giovanni Paolo II, 132, I-84084 Fisciano
  (Sa), Italy}
\author{S. De Palo}
\affiliation{CNR-IOM-Democritos National Simulation Centre, UDS Via Bonomea 265, I-34136, Trieste, Italy}
\affiliation{
Dipartimento di Fisica Teorica, Universit\`a Trieste, Strada Costiera 11, I-34014 Trieste, Italy}
\author{M. Di Dio}
\affiliation{CNR-IOM-Democritos National Simulation Centre, UDS Via Bonomea 265, I-34136, Trieste, Italy}
\author{E. Orignac}
\affiliation{Univ Lyon, Ens de Lyon, Univ Claude Bernard, CNRS, Laboratoire de Physique, F-69342 Lyon, France}

\begin{abstract}
  We consider a two leg bosonic ladder in a $U(1)$ gauge field with
  both interleg hopping and interleg repulsion. As a function of the flux, the interleg interaction
  converts the commensurate-incommensurate transition from the Meissner to a Vortex phase, into an Ising-type of transition towards a density wave phase.  A disorder point is also found after which the correlation functions develop a damped sinusoid behavior signaling a
  melting of the vortex phase. We discuss the differences on the phase diagram for attractive and repulsive interleg interaction.  In particular, we show how repulsion
  favors the Meissner phase at low-flux and a phase with a second incommensuration in the correlation functions for intermediate flux, leading to a
  richer phase diagram than in the case of interleg attraction. The effect of the temperature on the chiral current is also discussed.
\end{abstract}

\date{\today}

\maketitle

\section{introduction}
\label{sec:intro}
Trapped ultracold atoms have provided experimentalists with a unique ability
to realize highly tunable quantum simulators of many-body
model Hamiltonians,\cite{jaksch05_coldatoms,lewenstein07_coldatoms_review,bloch08_manybody}
including quasi-one dimensional systems.\cite{cazalilla2011}
Moreover, it has recently become possible to simulate the effect of an applied
magnetic field using two-photon Raman
transitions\cite{lin2011_soc,dalibard2011gauge,galitski2013_soc},
spin-orbit coupling\cite{celi2014} or
optical clock transitions\cite{Livi}. Such situation gives access to
a regime  where the interplay of low dimensionality, interaction and
magnetic field generates exotic phases such as bosonic analogues of the Fractional Quantum Hall
Effect.\cite{regnault2003_qhe}
The simplest system to observe nontrivial effects of an artificial
gauge field is the bosonic two-leg ladder.\cite{atala2014} Originally,
such systems were considered in the context of Josephson
junction arrays in magnetic
field\cite{kardar_josephson_ladder,orignac01_meissner,cha2011} and a
commensurate-incommensurate (C-IC) phase transition between a Meissner-like
phase with currents along the legs and a Vortex-like phase with
quasi long range ordered current loops was predicted.
However, in
Josephson junction systems, ohmic dissipation\cite{ambegaokar82_josephson_dissipation,korshunov_dissipative_josephson1d}
spoiled the quantum coherence required to observe such
transition. In cold atom systems, the Meissner and Vortex states have
been observed in a non-interacting case.\cite{atala2014}
Moreover, recent progress in superconducting qbits\cite{roushan2017} engineering
offer another promising path\cite{lehur2015,romero2017} for realization of low dimensional bosons
in artificial flux.
The availability of experimental systems has thus
renewed theoretical interest in the two leg bosonic ladder in a
flux\cite{dhar2012,dhar2013,petrescu2013,po2014,xu2014,zhao2014,wei2014,huegel2014,piraud2014,piraud2014b,keles2015,greschner2015,greschner2016,petrescu2015,barbiero2014,peotta2014,our_2015,our_2016,petrescu2016,barbarino2016,strinati2017,tokuno2014,uchino2015,bilitewski2016,greschner2017,guo2017,richaud2017,romen2017}.
These works have revealed  in that deceptively simple model  a zoo of ground state phases besides
Meissner-like and Vortex-like ones.
At commensurate filling,
Mott-Meissner and Mott-Vortex phases\cite{piraud2014b} as
well as chiral Mott
insulating phases\cite{dhar2012,dhar2013,petrescu2013,romen2017} have been
predicted. Meanwhile, with strong
repulsion and a flux $\Phi=2\pi n$ with $n$ the number of particles
per rung, bosonic analogs of the Laughlin states\cite{laughlin_FQHE}
are expected \cite{petrescu2015,petrescu2016,strinati2017}.
Interactions also affect the C-IC transition between the Meissner-like
and the Vortex-like phase\cite{natu2015}.
In a previous work\cite{orignac2017}, we have considered the effect of attractive interchain
interactions on the C-IC transition. Using an
analogy with statistical
mechanics of classical elastic systems on periodic substrates\cite{bohr1982,bohr1982b,haldane83_cic,schulz83_cic_vortices,horowitz_renormalization_incommensurable},
we showed that interchain attraction split the single
commensurate-incommensurate (C-IC) transition point into (a) an Ising transition
point between the Meissner-like phase and a density-wave phase, (b) a
disorder point\cite{stephenson1970a,stephenson1970b} where
incommensuration develops inside the density-wave phase, and (c) a
Berezinskii-Kosterlitz-Thouless (BKT)
transition\cite{berezinskii_2dxy,kosterlitz_thouless} where the density wave with incommensuration turns into the
Vortex-like phase. The density wave phase with
incommensuration can be identified as a melted vortex state while the transition (c) can be seen as a
melting of the vortex phase. The density
wave competing with the Meissner phase at Ising point (a) is induced by
interchain interaction\cite{cazalilla03_mixture,mathey08_supersolid,hu_pra_2009}
even in the absence of flux. We have verified
the existence of those phases in DMRG simulations of hard core
bosons.\cite{orignac2017} Since the analogy with classical elastic
systems holds irrespective of
the sign of the interchain interaction, a similar splitting of the
C-IC point should also in the repulsive case.
In the present manuscript, we show
that the splitting, if present, must occur in a much narrower region
of flux than in the attractive case.

The paper is organized as follows: In Section \ref{sec:model} we introduce the model and its bosonized version, here we also introduce the observables and their correlation functions. In Section \ref{sec:ising} we discuss the Ising transition and the disorder point by using a fermionization approach based on the Majorana fermion representation. Here we also briefly discuss the effect of the temperature on the spin current and momentum distribution.
In Section \ref{sec:2nd-ic} we discuss the emergence of the second incommensuration by using a unitary transformation approach and non-abelian bosonization.
Section \ref{sec:hard-core} presents the numerical results for the hard-core limit in the legs. In Section \ref{sec:ccl} we discuss the major results and give some conclusions.


\section{Model}
\label{sec:model}

We consider a model of bosons on a two-leg ladder in the presence of an artificial U(1) gauge field\cite{barbarino2016,strinati2017}:
\begin{eqnarray}
  \label{eq:full-lattice-ham}
  H=-t \sum_{j,\sigma} (b^\dagger_{j,\sigma} e^{i \lambda \sigma}
  b_{j+1,\sigma} +b^\dagger_{j+1,\sigma}  e^{-i \lambda \sigma}
  b_{j,\sigma}) \nonumber \\
  +\frac \Omega 2 \sum_{j,\alpha,\beta} b^\dagger_{j,\alpha}
  (\sigma^x)_{\alpha\beta} b_{j,\beta}
  + \sum_{j,\alpha,\beta} U_{\alpha \beta} n_{j\alpha} n_{j\beta},
\end{eqnarray}
where $\sigma=\uparrow,\downarrow$ represents the leg index,  $b_{j,\sigma}$ annihilates a boson on leg
$\sigma$ on the $j-$th site, $n_{j\alpha} = b^\dagger_{j\alpha}
b_{j\alpha}$, $t$ is the hopping amplitude along the chain, $\Omega$ is the tunneling between the legs,
$\lambda$ is the Peierls phase of the effective magnetic field associated to the gauge field, $U_{\uparrow\uparrow}=U_{\downarrow\downarrow}$ is the
repulsion between bosons on the same leg, $U_{\downarrow\uparrow}=U_\perp$ the
interaction between bosons on opposite legs. This model can be
mapped to a spin-1/2 bosons with spin-orbit interaction model\cite{our_2016}, where $\Omega$
is the transverse magnetic field, $\lambda$ measures the spin-orbit
coupling, $U_{\uparrow\uparrow}=U_{\downarrow\downarrow}$ is the
repulsion between bosons of identical spins, $U_{\downarrow\uparrow}=U_\perp$ the
interaction between bosons of opposite spins.


\subsection{Bosonized description}
\label{sec:bosonization}

Let us derive the low-energy effective theory for the
Hamiltonian~(\ref{eq:full-lattice-ham}), treating $\Omega$ and $U_\perp$ as
perturbations, and using Haldane's bosonization of interacting
bosons.\cite{haldane_bosons}

Introducing\cite{haldane_bosons} the fields $\phi_{\alpha}(x)$ and
$\Pi_\alpha(x)$ satisfying canonical commutation relations
$[\phi_\alpha(x),\Pi_\beta(y)]=i\delta(x-y)$ as well as the dual $\theta_\alpha(x) =\pi \int^x dy
\Pi_\alpha(y)$  of
$\phi_\alpha(x)$,  we can represent the boson annihilation operators as:

\begin{equation}
\frac{b_{j\sigma}}{\sqrt{a}}=\psi_\sigma(x)=e^{i\theta_\sigma(x)} \sum_{m=0}^{+\infty} A^{(\sigma)}_m \cos
  (2m \phi_\sigma(x) -2m \pi \rho^{(0)}_\sigma x),
\end{equation}
and the density operators\cite{haldane_bosons} as:
\begin{eqnarray}
\nonumber
  \frac{n_{j\sigma}}{a}&=&\rho_\sigma(x)=\rho_\sigma^{(0)} -\frac 1 {\pi} \partial_x
  \phi_\sigma  \\
&+& \sum_{m=1}^\infty B_m^{(\sigma)} \cos  (2m
  \phi_\sigma(x) -2m \pi \rho^{(0)}_\sigma x).
  \label{eq:boson-density}
\end{eqnarray}

Here, we have introduced the lattice spacing $a$, while $A_m$ and $B_m$
are non-universal coefficients that depend on the microscopic details
of the model. For integrable models, these
coefficients have been determined from Bethe Ansatz
calculations\cite{lukyanov_xxz_asymptotics,ovchinnikov2004,shashi2012}
while for
non-integrable models, they can be determined from numerical
calculations of correlation functions.\cite{hikihara03_amplitude_xxz,bouillot2010}

Introducing the canonically conjugate linear combinations:
\begin{eqnarray}
  \label{eq:c-field-def}
  \phi_c=\frac 1 {\sqrt{2}} (\phi_{\uparrow} + \phi_{\downarrow}) \;\;
  \Pi_c=\frac 1 {\sqrt{2}} (\Pi_{\uparrow} + \Pi_{\downarrow}), \\
\label{eq:s-field-def}
  \phi_s=\frac 1 {\sqrt{2}} (\phi_{\uparrow} - \phi_{\downarrow}) \;\;
  \Pi_s=\frac 1 {\sqrt{2}} (\Pi_{\uparrow} - \Pi_{\downarrow}),
\end{eqnarray}
the bosonized Hamiltonian can be rewritten as $H=H_c+H_s$, where
\begin{eqnarray}
  \label{eq:bosonized-c}
  H_c=\int \frac{dx}{2\pi} \left[u_c K_c (\pi \Pi_c)^2 +\frac {u_c}
    {K_c} (\partial_x \phi_c)^2 \right]
\end{eqnarray}
 describes the total density fluctuations for incommensurate filling when umklapp terms are irrelevant, and
\begin{eqnarray}
\nonumber
&&H_s = \int \frac{dx}{2\pi} \left[u_s K_s \left(\pi \Pi_s +\frac{\lambda}
{a \sqrt{2}} \right)^2 +\frac {u_s}
{K_s} (\partial_x \phi_s)^2 \right] \\
&&- 2 \Omega A_0^2 \int dx \cos
\sqrt{2} \theta_s + \frac{U_\perp a B_1^2} 2 \int dx \cos \sqrt{8} \phi_s
\label{eq:bosonized}
\end{eqnarray}
describes the antisymmetric density fluctuations. In
Eq.~(\ref{eq:bosonized}) and ~(\ref{eq:bosonized-c}),  $u_s$ and $u_c$ are respectively  the
velocity of antisymmetric and total density excitations, $A_0$ and $B_1$ are non universal coefficients\cite{giamarchi_book_1d}
while
$K_s$  and $K_c$ are  corresponding the Tomonaga-Luttinger (TL)
exponents. They can be expressed as a function of the velocity of
excitations $u$, and  Tomonaga-Luttinger
liquid exponent $K$ of the isolated chain as:
\begin{eqnarray}
  \label{eq:exponents-cs}
  u_c=u\left(1+\frac{U_\perp K a}{\pi u}\right)^{1/2} \\
u_s=u\left(1-\frac{U_\perp K a}{\pi u}\right)^{1/2} \\
 K_c=K \left(1+\frac{U_\perp K a}{\pi u}\right)^{-1/2}\\
  K_s=K \left(1-\frac{U_\perp K a}{\pi u}\right)^{-1/2}
\end{eqnarray}

For an isolated chain of hard core bosons,  we have $u=2t \sin (\pi
\rho^{0}_\sigma)$ and  $K=1$.
Physical observables can also be represented in bosonization.
The rung current, or the flow of bosons from the  upper leg to
the lower leg, is:
\begin{eqnarray}
\nonumber
  J_{\perp}(j)&=&-i \Omega (b^\dagger_{j,\uparrow} b_{j_\downarrow} -
  b^\dagger_{j,\downarrow} b_{j_\uparrow}). \\
&=& 2 \Omega A_0^2 \sin \sqrt{2} \theta_s +\ldots
\end{eqnarray}
The chiral current, {\it i.e.} the difference between the currents of
upper and lower leg, is defined as
\begin{eqnarray}
\label{eq:spin-current}
J_\parallel(j,\lambda) &=& -it \sum_{\sigma}\sigma (b^\dagger_{j,\sigma} e^{i \lambda \sigma}
b_{j+1,\sigma} - b^\dagger_{j+1,\sigma}  e^{-i \lambda  \sigma}
b_{j,\sigma}), \\
&=& \frac{u_s K_s}{\pi\sqrt{2}} \left(\partial_x \theta_s +\frac{\lambda}{a \sqrt{2}} \right).
\end{eqnarray}
The  density difference between the chains $S_j^z=n_{j\uparrow}
-n_{j\downarrow}$, is written in bosonization as:
\begin{eqnarray}
  S_j^z=-\frac{\sqrt{2}}{\pi} \partial_x \phi_s -2 B_1 \sin (\sqrt{2}
  \phi_c -\pi \rho x) \sin \sqrt{2} \phi_s,
\end{eqnarray}
while the density of particles per rung is:
\begin{eqnarray}
  n_j=-\frac{\sqrt{2}}{\pi} \partial_x \phi_c -2 B_1 \cos (\sqrt{2}
  \phi_c -\pi \rho x) \cos \sqrt{2} \phi_s.
\end{eqnarray}

Let us discuss some simple limits of the Hamiltonian~(\ref{eq:bosonized}).
When $\Omega\ne 0$, $U_\perp=0$, and $\lambda\rightarrow 0$, the antisymmetric
modes Hamiltonian Eq.~(\ref{eq:bosonized}) reduces to a quantum
sine-Gordon Hamiltonian.
For $K_s>1/4$, the spectrum of $H_s$ is gapped and the system is in the so-called Meissner state\cite{kardar_josephson_ladder,orignac01_meissner} characterized
by $\langle \theta_s\rangle =0$.
In such state, the chiral current increases linearly with the applied flux at
small $\lambda$, while the average rung current $\langle J_\perp
\rangle =0$  and its correlations $\langle  J_{\perp}(j)
J_{\perp}(0)\rangle$ decay exponentially with distance. The
antisymmetric density
correlations also decay exponentially with distance, while
the symmetric ones behave as:
\begin{eqnarray}
  \langle n_i n_j \rangle = -\frac{2 K_c}{\pi^2 (i-j)^2} + e^{-|i-j|/\xi}
  \frac{\cos \pi n (i-j)}{|i-j|^{K_c}},
\end{eqnarray}
 where $\xi$ is the correlation length resulting from the spectral gap
of $H_s$.
With $\Omega=0,U_\perp \ne 0$, the antisymmetric density fluctuations
Hamiltonian~(\ref{eq:bosonized}) becomes again a quantum sine-Gordon
model that can be related to the previous one by the duality
transformation $\theta_s \to 2 \phi_s, \phi_s \to \theta_s/2, K_s \to
1/(4K_s)$.  For $K_s<1$, the Hamiltonian $H_s$ has a gapped spectrum
and
$\langle \phi_s \rangle =
\frac{\pi}{\sqrt{8}}$ for $U_\perp>0$ yielding a zig-zag density wave
ground state and $\langle \phi_s \rangle =0$
for $U_\perp<0$ yielding a rung density wave ground
state.\cite{orignac98_vortices,cazalilla03_mixture,mathey07_mixture,mathey07_bose_fermi,mathey_pra_2009,hu_pra_2009}
In both density wave states, the expectation values of the spin and
conversion current vanish, and their correlations decay
exponentially. The Green's functions of the bosons also decay
exponentially, so that the momentum distribution only has
a Lorentzian shaped maximum at $k=0$.
However, in the zig-zag density wave state ($U_\perp>0)$, we have:
\begin{eqnarray}
  \langle S_j^z  S_k^z \rangle& \sim& C_1 e^{-|j-k|/\xi} + C_2 \frac{\cos \pi
    n(j-k)}{|j-k|^{K_c}}, \\
  \langle n_j  n_k \rangle& \sim& - \frac{2 K_c}{\pi^2 (j-k)^2} + C_3 \frac{\cos \pi
    n(j-k)}{|j-k|^{K_c}}  e^{-|j-k|/\xi} ,
\end{eqnarray}
while in the rung density wave ($U_\perp <0$),
\begin{eqnarray}
   \langle n_j  n_k \rangle & \sim& -\frac{2 K_c}{\pi^2 (j-k)^2} + C'_3 \frac{\cos \pi
    n(j-k)}{|j-k|^{K_c}}, \\
   \langle S_j^z  S_k^z \rangle & \sim& C'_1 e^{-|j-k|/\xi} + C'_2 \frac{\cos \pi
    n(j-k)}{|j-k|^{K_c}}  e^{-|j-k|/\xi}
\end{eqnarray}
where $K_c$ depends on the interleg interaction, increasing when it is
attractive and decreasing when it is repulsive as indicated in
Eq.~(\ref{eq:exponents-cs}). The behavior of density correlations in
real space is reflected in the corresponding static structure factors:
\begin{eqnarray}
  \label{eq:strucfac-def}
  S^c(q)=\sum_j e^{-iqj} \langle n_j n_0\rangle, \\
  S^s(q)=\sum_j \langle S_j^z  S_0^z \rangle.
\end{eqnarray}
In all phases, $S^c(q\to 0) =\frac{2 K_c}{\pi} |q| +o(q)$, while $S_s(q)\sim
S_s(0)+A q^2 + o(q^2)$  indicating that symmetric excitations are always
gapless while antisymmetric excitations are always gapped. However, in
the rung density
wave, $S_c(q \to \pi n)$ has a power law divergence $\sim |q-\pi n|^{K_c-1}$ (if
$K_c<1$) or a cusp $\sim C+C' |q-\pi n|^{K_c-1}$ (if $1<K_c<2$)  and
$S_s(q\to \pi n)$ has only a Lorentzian-shaped maximum while in the
zig-zag density-wave, $S_s(q\to \pi n)$ shows a cusp or singularity while
$S_c(q\to \pi n)$ has a Lorentzian-shaped maximum.
The case of $U_{\alpha\alpha}=+\infty$ is peculiar as $K_s \to 1$. The
Hamiltonian~(\ref{eq:full-lattice-ham})  can then be mapped to the
Fermi-Hubbard model (see Sec.~\ref{sec:hubbard}). Bosonization of the
Fermi-Hubbard
model\cite{giamarchi_book_1d}  shows
that the operator $\cos \sqrt{8} \phi_s$ is marginal in the
renormalization group
sense. On the attractive side,\cite{giamarchi_book_1d}
it is marginally relevant, and the density wave exists for all
$U_\perp<0$.  However, on the
repulsive side,  $\cos \sqrt{8} \phi_s$ is marginally irrelevant and
the staggered density wave is absent.

With both $\Omega$ and $U_\perp$ nonzero and $\lambda=0$, the Hamiltonian $H_s$ becomes the
self-dual sine-Gordon model.\cite{jose_planar_2d,lecheminant2002sdsg}
When both cosines are relevant (\textit{i. e.} $1/4<K_s<1$) the
Meissner phase (stable for $|\Omega| \gg |U_\perp|$)  is competing
with the density wave phases (stable in the opposite limit).
The competing phases are separated by an Ising critical
point.\cite{jose_planar_2d,lecheminant2002sdsg}
In the case of $U_{\alpha\alpha}=+\infty$, since the density wave
is absent for $U_\perp>0$, one only has the Meissner state for all
$U_\perp>0$.
By contrast, for $U_\perp<0$, the charge density wave exists at
$\Omega=0$ and an Ising critical point is present. Thus, phase
diagrams for $U_\perp>0$ and $U_\perp<0$ are very different.

In the presence of flux ($\lambda \ne 0$), the density wave phases are
stable.
However,for $U_\perp=0$,  in the Meissner phase,\cite{kardar_josephson_ladder,orignac01_meissner} when the flux $\lambda$ exceeds the threshold $\lambda_c$
the commensurate-incommensurate transition takes place:
\cite{japaridze_cic_transition,pokrovsky_talapov_prl,schulz_cic2d}
the ground state of $H_s$ then presents a non-zero density of
sine-Gordon solitons forming a Tomonaga-Luttinger liquid.\cite{kardar_josephson_ladder,orignac01_meissner}
The low energy properties of the incommensurate phase are described by the effective Hamiltonian:
\begin{equation}
\label{eq:eff-cic}
  H^*= \int \frac{dx}{2\pi} \left[ u^*_s(\lambda) K^*_s(\lambda) (\pi \Pi^*_s)^2 +\frac {u^*_s(\lambda)} {K^*_s(\lambda)} (\partial_x \phi)^2 \right],
\end{equation}
where $\Pi_s = \Pi_s^* + \langle \Pi_s\rangle(\lambda)$. Near the
transition point $\lambda_c$, $ \langle \Pi_s\rangle(\lambda) \sim C
\sqrt{\lambda-\lambda_c}$. Moreover, as $\lambda \to \lambda_c+0$,
$K_s^*(\lambda)$ goes to a limiting value $K_s^{(0)}$ such
that\cite{schulz_cic2d,chitra_spinchains_field} the scaling dimension
of $\cos \sqrt{2} \theta_s$ becomes $1$. Since the scaling dimension
of $\cos \sqrt{2} \theta_s$ with a Hamiltonian of the
form~(\ref{eq:eff-cic}) is $1/[2K_s^*(\lambda)]$ one finds
$K_s^{(0)}=1/2$.
In that incommensurate phase, called the Vortex
state\cite{orignac01_meissner} in the ladder language, $\langle
J_\parallel(j)\rangle$
decreases and eventually vanishes for large flux values. Meanwhile,
the conversion current correlations, density correlations
 and the Green's functions of the
bosons decay with distance as a power law damped sinusoids.
The effect of the interaction between identical spins on the commensurate-incommensurate transition has been largely investigated
both numerically and theoretically.\cite{petrescu2013,piraud2014,piraud2014b,wei2014,our_2016}

Since $U_\perp$ can give rise a phase competing with the Meissner
state in the absence of flux,
its effect on the commensurate incommensurate transition
induced by $\lambda$ needs to be considered.
Indeed, near the transition the scaling dimension of the field $\cos\sqrt{8} \phi_s$ is
$2 K_s^*(\lambda)\simeq 1$, thus the $\cos \sqrt{8} \phi_s$ term in Eq.~(\ref{eq:bosonized}) is
relevant and causes a gap opening.\cite{horowitz_renormalization_incommensurable,haldane83_cic}
A fermionization approach\cite{bohr1982,bohr1982b} allows to show that the flux induced
transition remains in the Ising universality class.
Moreover, this approach also predicts the existence of a disorder point\cite{stephenson1970a,stephenson1970b} where incommensuration develops in some correlation functions even though
the gap and the density wave phase persist.
For instance, the bosonic Green function reads:
\begin{eqnarray}
\nonumber
  \langle b_{j,\sigma} b^\dagger_{k,\sigma} \rangle &=& \langle e^{i \frac{\theta_c (ja)}{\sqrt{2}}}  e^{-i \frac{\theta_c (ka)}{\sqrt{2}}}\rangle \langle   e^{i \sigma \frac{\theta^*_s (ja)}{\sqrt{2}}}  e^{-i \sigma \frac{\theta^*_s (ka)}{\sqrt{2}}}\rangle \\
  &\sim& \left(\frac 1 {|j-k|}\right)^{\frac 1 {4K_c}} e^{i \sigma q(\lambda)(j-k)} e^{-|j-k|/\xi},
\end{eqnarray}
where $q=\pi \langle \Pi_s \rangle/(a\sqrt{2})$ and consequently
the momentum distribution \begin{eqnarray}
  n_\sigma(k) = \sum_j \langle b_{j,\sigma}^\dagger b_{0,\sigma}
  \rangle e^{-ikj},
\end{eqnarray}
instead of showing
power-law divergences\cite{our_2015} at momentum $\pm q$ as in the
vortex state, presents Lorentzian-shaped maximas.
In the bosonization picture, the disorder point can be
understood as the superposition of the incommensuration induced by
$\lambda \Pi_s$ and the gap opened by $\cos \sqrt{8} \phi_s$.
As $\lambda$ further increases, the dimension $K_s^*(\lambda)$ recovers the value
$K_s$. In the case of $K_s>1$, there is a second critical point,
$\lambda=\lambda_{BKT}$ where $K_s^*(\lambda_{BKT})=1$ and the $\cos
\sqrt{8} \phi_s$ operator becomes marginal. At that point, a
Berezinskii-Kosterlitz-Thouless\cite{berezinskii_2dxy,kosterlitz_thouless}
takes place,\cite{haldane83_cic} from the density wave phase to the
gapless  vortex
state.\cite{orignac_vortex_ladder} This allows to interpret the
density wave state with incommensuration as a melted vortex state.
By contrast, if $K_s<1$, the ground
state remains in a gapped density wave for all values of
$\lambda>\lambda_c$.

\section{Ising transition and disorder point}
\label{sec:ising}

As discussed above in Sec. \ref{sec:bosonization}
the application of the flux gives rise to an Ising transition point
followed by a disorder point both of which can be described using a Majorana
fermion representation.

\subsection{Majorana Fermions representation and Quantum Ising transition}\label{sec:majoranas}

Let us now consider a value of the flux close at the commensurate-incommensurate transition, when $K_s=1/2$, fermionization\cite{bohr1982,bohr1982b} leads
to a  a detailed picture of the transition between the Meissner state and
the density wave states. The fermionized Hamiltonian reads\cite{orignac2017}:
\begin{eqnarray}
\nonumber
H&=&-i\frac {u_s} 2 \int dx \sum_{j=1}^2 (\zeta_{R,j} \partial_x
  \zeta_{R,j} - \zeta_{L,j} \partial_x \zeta_{L,j}) \\
\nonumber
&&-i \sum_{j=1,2}  m_j \int
  dx  \zeta_{R,j} \zeta_{L,j}  \\
&&-ih  \int dx  (\zeta_{R,1} \zeta_{R,2}
  +  \zeta_{L,1} \zeta_{L,2}) + \int dx \frac{h^2}{2\pi u_s}
  \label{eq:ham-majorana}
\end{eqnarray}
where $m_j=m+(-)^{j-1}\Delta$ with,
\begin{eqnarray}
  \label{eq:ham-dirac-params-h}
  h&=& -\frac{ \lambda u_s K_s}{a}, \\
  \label{eq:ham-dirac-params-m}
  m&=&2\pi \Omega A_0^2 a, \\
  \label{eq:ham-dirac-params-delta}
  \Delta&=&\frac \pi 2 U_\perp (B_1 a)^2,
\end{eqnarray}
and $\{\zeta_{\nu,j}(x),\zeta_{\nu',j'}(x')\}=\delta_{nu,\nu'}
\delta_{j,j'} \delta(x-x')$ are Majorana fermion field operators.

Hamiltonians of the form (\ref{eq:ham-majorana}) have previously been
studied in the context of spin-1 chains in magnetic
field\cite{tsvelik_field,wang2003field,essler04_spin1_field} or
spin-1/2 ladders\cite{shelton_spin_ladders,nersesyan_biquad} with
anisotropic interactions.\cite{citro02_dm_ladders}


The eigenvalues of (\ref{eq:ham-majorana}) are:
\begin{eqnarray}
\nonumber
E_\pm(k)^2&=&(u_s k)^2 + m^2 + h^2 + \Delta^2  \\
&& \pm 2 \sqrt{h^2 (u_s k)^2 + h^2 m^2 +\Delta^2 m^2}.
  \label{eq:eigenenergies}
\end{eqnarray}
For $m=\sqrt{h^2+\Delta^2}$, $E_-(k)=u \frac \Delta m |k| + O(k^2)$,
and a single Majorana fermion mode becomes massless at the transition\cite{bohr1982} between the Meissner and the
density wave state as expected at an
Ising\cite{mccoy_revue_qft} transition.
As a consequence, at the transition, the Von Neumann entanglement entropy $S_{vN}=\frac 1 3 (c_{c}+c_{Ising})\ln L
= \frac 1 2 (1 + 1/2) \ln L$, while away from the transition it is
$S_{vN}=\frac 1 3 c_c \ln L=\frac 1 3 \ln L$ since the total density
modes $\phi_c$ are always gapless. A more detailed discussion of finite size
scaling of entanglement entropies is found in
Ref.~\onlinecite{campostrini2014}.

\subsection{Ising order and disorder parameters}

At the point $K_s=1/2$,
the bosonization operators $\cos \theta_s/\sqrt{2}$, $\sin \theta_s/\sqrt{2}$, $\cos \sqrt{2}\phi_s$ and
$\sin \sqrt{2} \phi_s$ can be expressed in terms of the Ising order and disorder
operators associated with the Majorana fermions operators of  Eq.~(\ref{eq:ham-majorana})
as:\cite{zuber_77,schroer_ising,boyanovsky_ising,nersesyan2001_ising}
\begin{eqnarray}
  \label{eq:double-ising}
  \cos \frac{\theta_s}{\sqrt{2}} = \mu_1 \mu_2 \; \sin \frac{\theta_s}{\sqrt{2}} = \sigma_1 \sigma_2, \\
 \cos \sqrt{2}\phi_s = \sigma_1 \mu_2 \; \sin \sqrt{2}\phi_s = \mu_1
 \sigma_2.
\end{eqnarray}
With our conventions, for $m_j>0$ we have $\langle \mu_j \rangle \ne 0, \langle \sigma_j \rangle =0$
while  $m_j<0$ we have $\langle \mu_j \rangle = 0, \langle \sigma_j
\rangle \ne  0$.
In terms of the Ising order and disorder fields,
\begin{eqnarray}
  \label{eq:ising-mapping-Upos}
 && b_{j,\sigma} =e^{i\frac{\theta_c}{\sqrt{2}}} (\mu_1
  \mu_2 + i \mathrm{sign}(\sigma) \sigma_1 \sigma_2) \\
&&  S_j^z = i(\zeta_{R,1} \zeta_{R,2} -\zeta_{L,1} \zeta_{L,2})
\nonumber \\ &&  -2 B_1 \sin (\sqrt{2}
  \phi_c -\pi \rho x) \mu_1 \sigma_2, \\
 &&  n_j = -\frac{\sqrt{2}} \pi \partial_x \phi_c  -2 B_1 \sin (\sqrt{2}
  \phi_c -\pi \rho x) \mu_2 \sigma_1,
\end{eqnarray}

Let's consider first the case of $h=0$, $\Omega>0$.
For $U_\perp=0$ the system  is in the Meissner phase
with $\langle \mu_1 \rangle \langle \mu_2 \rangle
\ne 0$.
As $U_\perp>0$ increases, $m_2=m-\Delta$ changes sign, so that
$\langle \sigma_2 \rangle \ne 0$ while $m_1$ remains positive and
$\mu_1\ne 0$. As a result, $\langle \sin \sqrt{2} \phi_s\rangle \ne
0$, and we recover the zig-zag density wave phase.\cite{natu2015}
With $U_\perp<0$, $m_2$ remains positive, while $m_1$ is changing
sign. As a result, for large $|U_\perp|$,
$\langle \sigma_1\rangle \ne 0$ giving a  nonzero $\langle \cos
\sqrt{2} \phi_s\rangle$
and a rung density wave sets in.

Instead as a function of $h$,
we stress that in case of fixed $U_\perp,\Omega$ and variable $h$, a
phase transition is possible only if $m^2-\Delta^2>0$, \textit{i. e.}
only when for $h=0$ we have $\langle \mu_1 \rangle \langle \mu_2
\rangle \ne 0$.
Then, for $h>\sqrt{m^2-\Delta^2}$,  we will have $\langle
\mu_1\rangle \langle \sigma_2\rangle \ne 0$ (for $U_\perp>0$) or  $\langle
\mu_2\rangle \langle \sigma_1 \rangle \ne 0$ (for $U_\perp<0$).
Therefore, as in the case of
the transition as a function of $U_\perp$, one of the pairs of dual Ising
variable is becoming critical at the transition while the other
remains spectator.

\subsection{Disorder point}
\label{greens_function}

  The correlators of the Majorana fermion operators $\langle \zeta_{\nu,j}(x)
  \zeta_{\nu',j'}(x')\rangle $  can be
obtained from just two integrals\cite{orignac2017}:
\begin{eqnarray}
  \label{eq:integral-RS}
  I_1(x)=\int \frac{dk}{2\pi} \frac{e^{ik x}}{E_+(k) E_-(k) (E_+(k) +
    E_-(k))}, \\
 I_2(x)=\int \frac{dk}{2\pi} \frac{e^{ik x}}{(E_+(k) +
    E_-(k))}
\end{eqnarray}
by taking the appropriate number of derivatives with respect to $x$.

To estimate the asymptotic behavior of the Green's functions, one can
apply a contour integral method\cite{bender78_book} as detailed in the
Appendix. The long distance behavior is determined by
 the branch cut singularities of the denominators in the upper
half plane. For $I_2$, the cut is obtained for $u k=\pm i
\sqrt{m^2(1+\Delta^2/h^2)} \cosh \phi$, so
$I_2(x)=O(e^{-|x|\sqrt{m^2(1+\Delta^2/h^2)/u}})$. As a result, the long
  distance behavior is dominated by $I_1(x)$.
For $h<m$, its branch cut extends along the imaginary
axis from $i|\Delta -\sqrt{m^2-h^2}|/u< k < i(\Delta
+\sqrt{m^2-h^2})$, giving  $I_1(x) \sim e^{-\frac{|\Delta-\sqrt{m^2-h^2}||x|}
  u}$. This recovers the correlation length diverging as $\sim
|m-\sqrt{h^2+\Delta^2}|^{-1}$ near the Ising transition.

For $h>m$, the denominator in $I_1$ has two branch cuts that terminate into two branch points. The long distance behavior of
$\bar{I}_1$ is determined by these two branch points as:
\begin{equation}
  \label{eq:disorder-i1bar}
  \bar{I}_1(x) \sim e^{-\frac{\Delta |x|} u} \left[e^{i
      \frac{\sqrt{h^2-m^2}|x|} u} \varphi_1(x) + e^{-i
      \frac{\sqrt{h^2-m^2}|x|} u} \varphi_1(x)^*\right],
\end{equation}
with $|\varphi_1(x)|=O(x^{-1/2})$,  so that oscillations of wavevector $\sqrt{h^2-m^2}/u$ appear in the
real space Majorana fermion correlators for $h>m$. The point
$h=m$ is called a disorder
point.\cite{stephenson1970a,stephenson1970b}

If we calculate equal time correlation functions of the conversion
current using Wick's theorem, the result depends  on
 products of two Green's functions. The conversion current thus shows
exponentially damped oscillations with wavevector $2\sqrt{h^2-m^2}/u_s$ and correlation length $u_s/(2\Delta)$.

Moreover, the correlation functions of the Ising order and disorder
fields are expressed in
terms of Pfaffians of antisymmetric matrices whose elements are
expressed in terms of the Majorana fermion Green's
functions.\cite{mccoy_revue_qft} The presence of exponentially damped
oscillations in the Majorana fermions Green's function thus also
affects correlation functions of Ising order and
disorder operators.\cite{wang2003field} More precisely, when the large
flux ground state is the CDW, for long distances:
\begin{eqnarray}
  \langle \sigma_1(x) \mu_2(x) \sigma_1(0) \mu_2(0)\rangle& \sim&
  \frac{e^{-\frac{2\Delta|x|} u}}{r}, \\
   \langle  \mu_1(x) \sigma_2(x)  \mu_1(0) \sigma_2(0)\rangle& \sim&
   (\langle \mu_1 \sigma_2 \rangle)^2 \ne 0,
   \\
 \langle  \mu_1(x) \mu_2(x) \mu_1(0) \mu_2(0) \rangle &\sim&
 \frac{e^{-\frac{\Delta|x|} u}}{\sqrt{r}} \times \nonumber \\ &&\times
 \cos \left(\frac{\sqrt{h^2-m^2} x}
   u\right),
 \end{eqnarray}
and when the ground state is the zig-zag density wave the long distance correlations of $\sigma_2
\mu_1$ and $\sigma_1 \mu_2$ are exchanged.

\subsection{Effect of finite temperature}
\label{sec:free-energy}

\begin{figure}[ht]
  \centering
  \includegraphics[width=9cm]{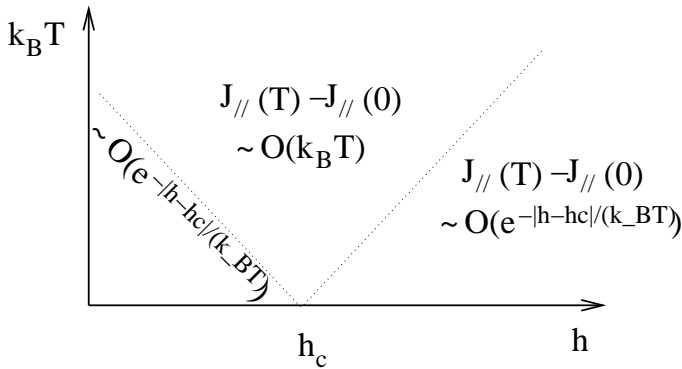}
  \caption{Crossover diagram for the current. Below the dashed line,
the low temperature region with $h<h_c$ is the ``renormalized classical''
regime, while the low temperature region with $h>h_c$ is the
disordered regime. In both of these regions, the finite temperature
correction to the zero temperature current is exponentially
small. Above the dashed line, in the quantum critical region,
thermal corrections are $O(k_B T)$.}
  \label{fig:crossover}
\end{figure}

From the eigenenergies~(\ref{eq:eigenenergies}), we find the free
energy per unit length as:
\begin{eqnarray}
  \label{eq:free-energy}
  f=\frac {F} L = \frac{h^2}{2\pi u_s} - k_B T \sum_{r=\pm} \int_0^\Lambda
  \frac{dk}{\pi} \ln\left[2 \cosh\left(\frac{E_r(k)}{2k_B T}
    \right)\right] \nonumber \\
\end{eqnarray}
The spin current is $J_s=-\frac{u_s K_s}a \partial_h f$ with:
\begin{eqnarray}
  \label{eq:current-Tpos}
  \frac{\partial f}{\partial h} = \frac{h}{\pi u_s} - \sum_{r=\pm} \int_0^\Lambda
  \frac{dk}{2\pi} \tanh \left(\frac{E_r(k)}{2k_B T}
    \right)\frac{\partial E_r(k)}{\partial h}
\end{eqnarray}
The integral (\ref{eq:current-Tpos}) is convergent in the limit
$\Lambda \to +\infty$.
We can split (\ref{eq:current-Tpos}) into a ground state contribution
and a thermal contribution:
\begin{eqnarray}
  \frac{\partial f}{\partial h} =  \frac{\partial e_{GS}}{\partial h}
  +  \sum_{r=\pm} \int_0^\Lambda
  \frac{dk}{\pi} \frac{2}{e^{\frac{E_r(k)}{k_B T}} +1} \frac{\partial
      E_r(k)}{\partial h},
\end{eqnarray}
and we see that away from the critical point, the latter contribution is
$O(e^{-E_-(0)/(k_B T)})$ when $E_-(0) \gg k_B T$. For $E_-(0) \ll k_B
T$ the thermal contribution becomes $O(k_B T)$. A crossover
diagram\cite{sondhi_qcp,sachdev_book} is represented on
Fig.\ref{fig:crossover}. The region where the corrections are linear
in temperature is the quantum critical region.

\begin{figure}
  \centering
  \includegraphics[width=9cm]{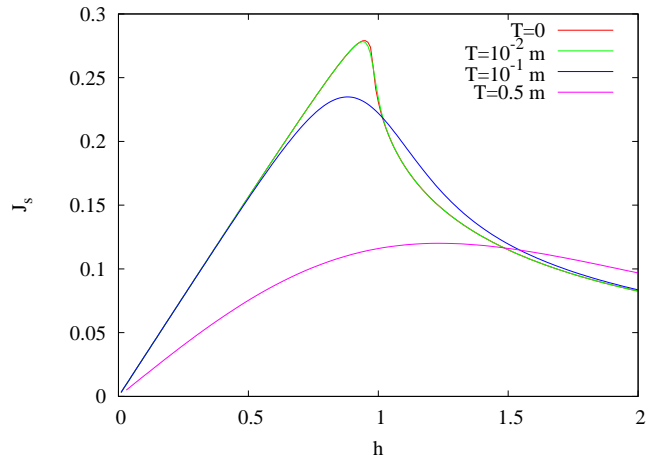}
  \caption{The current versus flux curves for $\Delta=0.2 m$ at
    varying temperature. For temperatures small ($T=0.01 m$) compared
    with $\Delta,m$ the curve is indistinguishable from the zero
    temperature curve. As temperature become comparable with $\Delta$
    the maximum of the current becomes broader and shifts to higher
    flux. For temperature comparable with $m$, the maximum becomes
    very broad.}
  \label{fig:varyingT}
\end{figure}

At fixed temperature, varying the applied flux, two regimes are
possible. For $k_B T \ll \mathrm{min}(\Delta, m)$, only a narrow region of flux
around the critical flux is inside the quantum critical region, and
the current versus flux curve is barely modified. For $k_B T \gg
\mathrm{min}(\Delta, m)$, the current versus flux curve is showing a
broadened maximum that shifts progressively to higher flux. This
behavior is shown on Fig.~\ref{fig:varyingT}.

If we turn to the current susceptibility, which has a logarithmic
divergence at the critical flux in the ground state, its positive
temperature expression is:

\begin{eqnarray}
  \label{eq:suscep-Tpos}
  \frac{\partial^2 f}{\partial^2 h} = && \frac 1 {\pi u} -\sum_{r=\pm} \int_0^{+\infty}
  \frac{dk}{2\pi} \left[\frac{\partial^2 E_r(k)}{\partial h^2} \tanh
    \left(\frac{E_r(k)}{2k_B T}\right) + \right. \nonumber \\
    && \left. \left(\frac{\partial
        E_r(k)}{\partial h}\right)^2 \frac 1 {2k_B T \cosh^2
      \left(\frac{E_r(k)}{2k_B T}\right)}\right],
\end{eqnarray}


Exactly at the critical point $h=\sqrt{m^2-\Delta^2}$, we find that:
\begin{equation}
  \frac{\partial^2 E_-(k)}{\partial h^2} = \frac{h^2}{2m \Delta u |k|}
  + O(|k|),
\end{equation}
so that $\frac{\partial J}{\partial h} \sim \frac{h^2}{4\pi m \Delta
  u} \ln (1/T)$.
In the general case, the divergence of $\partial_h J$ is controlled by
the integral:
\begin{eqnarray}
  \int_0^\Lambda \frac{dk}{E_-(k)} \tanh\left(\frac{E_-(k)}{2k_B
      T}\right).
\end{eqnarray}
If we take $T=0$, the integral will have a logarithmic
divergence in the limit of $h \to \sqrt{m^2 -\Delta^2}$, indicating
the Ising transition. However, for any finite $T$, the hyperbolic
tangent will cutoff the divergence for $E_-(k) \ll k_B T$, and give
instead a maximum scaling as $\sim \ln (1/T)$. Therefore, one expects
that $\partial_h J \sim - \ln [(m-\sqrt{h^2+\Delta^2})^2 + (k_B
T)^2]$. Thus, for very low temperature, the slope of the curve $J$
versus $h$ presents a maximum at $h=\sqrt{m^2-\Delta^2}$ indicating
the presence of an inflection point instead of the vertical tangent
obtained at $T=0$.
If we turn to correlation functions, since our system is one
dimensional, at any nonzero temperature its correlation functions
always decay
exponentially.\cite{landau-statmech-english} However, in the quantum Ising
chain, the correlation
length of operators that are long range ordered at zero temperature
has been found\cite{sachdev_ising,sachdev1997} to behave as $\sim
u_s (T M)^{-1/2} e^{M/T}$ where $M$ is the gap at zero
temperature. By contrast, operators with short-range ordered
correlations in the ground state still have a correlation length $\sim
u_s/M$. The difference between the two classes of operators
thus remain distinguishable until $T\sim M$. Therefore, in the
``renormalized classical'' region, the zero temperature power law
peaks in $n_\sigma(k \to 0)$ turns into a narrow Lorentzian maximum,
while the Lorentzian maximas in $S^{c/s}(k)$ and $C(k)$ remain broad.
The distinction between CDW and Meissner phase is lost only at a
temperature $k_B T \sim E_-(0)$.

\section{Second incommensuration with repulsive interaction}
\label{sec:2nd-ic}

In previous investigations\cite{our_2015,our_2016} a second
incommensuration (2IC) was obtained when the flux $\lambda=\pi n$ for a
two-leg ladder of hard core bosons. The 2IC can be associated to the interchain hopping and manifests
in the periodic oscillations of the correlation functions at wavevectors formed
by a linear combinations of $\lambda$ and $\Omega$.
A
very
simple picture of the second incommensuration can be obtained in the
limit $U_\perp \gg t$  where one can use  a Jordan-Wigner representation for the bosons. \\
Using a gauge
transformation, the Hamiltonian (\ref{eq:full-lattice-ham}) can be
rewritten:

\begin{eqnarray}
\nonumber
  H&=&-t \sum_{j,\sigma}  (b^\dagger_{j+1,\sigma} b_{j,\sigma} +
  b^\dagger_{j,\sigma} b_{j+1,\sigma})  + U\sum_j n_{j\uparrow}
   n_{j\downarrow} \\
&&+ \frac \Omega 2 \sum_{j} (e^{i
    \lambda j}
  b^\dagger_{j,\uparrow} b_{j,\downarrow}  + e^{-i
    \lambda j}
   b^\dagger_{j,\downarrow}  b_{j,\uparrow}).
  \label{eq:gauged-ham}
\end{eqnarray}

In terms of  the Jordan-Wigner fermions~(\ref{eq:jordan-wigner}) the interchain hopping has, in general,  a complicated non-local
expressions:
\begin{equation}
   b^\dagger_{j,\uparrow} b_{j,\downarrow} = c^\dagger_{j,\uparrow}
   \eta_{j\uparrow} \eta_{j\downarrow} c^\dagger_{j,\downarrow} e^{i
     \pi \sum_{k<j} (n_{k\uparrow} + n_{k\downarrow})},
\end{equation}

However, at half-filling, the charge is gapped so that one can
approximate,
\begin{equation}
  e^{i
     \pi \sum_{k<j} (n_{k\uparrow} + n_{k\downarrow})} \simeq (-)^j,
\end{equation}
 and the remaining
gapless spin mode described by an effective spin chain model:
\begin{equation}
  \label{eq:spinchain}
  H=\frac{4t^2}U \sum_n \vec{S}_n \cdot \vec{S}_{n+1}.
\end{equation}
The antihermitian operator $ \eta_{j\uparrow} \eta_{j\downarrow}$
commutes with the Hamiltonian, and can be replaced by one of its
eigenvalues $\pm i$. Then, the interchain hopping reduces to:
\begin{equation}
  \frac{\Omega} 2  \sum_j (e^{i(\lambda-\pi) j} i c^\dagger_{j_\uparrow}
  c_{j_\downarrow} + H. c.),
\end{equation}
and, having in mind the Jordan-Wigner transformation (\ref{eq:jordan-wigner}),
it reduces to  $\Omega \sum_j S_j^y$ when $\lambda=\pi$.
Therefore, it acts on the spin chain~(\ref{eq:spinchain})  as a uniform
magnetic field, and induces a magnetization along the $y$
axis. Such magnetization also gives rise to
incommensuration\cite{giamarchi_book_1d} in the correlation functions of the spin components
$x$ and $z$. This treatment represents the simplest way to understand the origin of a second-incommensuration in the correlation functions.
However, in the case away from half-filling, the second incommensuration could not be deduced
as straightforwardly\cite{our_2016} and one had to resort to a modified
mean-field theory. \\
Here, we want to present another approach, using a
canonical transformation that avoids some of the shortcomings of the
mean-field theory.
If we bosonize the Jordan-Wigner fermionic version of the Hamiltonian (\ref{eq:gauged-ham}) we obtain:
\begin{eqnarray}
&&  H=\sum_{\nu=c,s} \int \frac{dx}{2\pi} \left[u_\nu K_\nu (\pi \Pi_\nu)^2 +
  \frac{u_\nu} {K_\nu} (\partial_x \phi_\nu)^2\right] \nonumber \\
&& + \frac{\Omega}{2\pi a} \int dx \left[e^{i\sqrt{2} \phi_c} (e^{-i \sqrt{2} (\theta_s +
  \phi_s)} +  e^{-i \sqrt{2} (\theta_s -
  \phi_s)}) + \mathrm{H. c.} \right] \nonumber \\
&& -\frac{2 g_{1\perp}}{(2\pi a)^2} \int dx \cos \sqrt{8} \phi_s.
\end{eqnarray}
Then we consider the action of the unitary operator:
\begin{eqnarray}
  \label{eq:unitary-op}
  U=\exp\left[ -i \int \frac {dx} \pi \phi_c(x) \partial_x \phi_s\right]
\end{eqnarray}
Such canonical transformation gives a
controlled approximation in the limit of $K_c \to 0$. Indeed,
by rescaling $\phi_c \to \sqrt{K_c} \phi_c$, $\theta_c \to
\theta_c/\sqrt{K_c}$, the unitary transformation~(\ref{eq:unitary-op})
becomes close to identity as $K_c \ll 1$ and the resulting
perturbations in the transformed Hamiltonian are then small.
The transformed Hamiltonian is:
\begin{eqnarray}
  \label{eq:u-trasf-h}
  U^\dagger H U &=&\int \frac{dx}{2\pi} \left[u_c K_c (\pi \Pi_c)^2 + u_s K_s (\pi
    \Pi_s)^2 \right. \nonumber \\
    &+& \left. \left(u_s K_s + \frac{u_c}{K_c} \right) (\partial_x
    \phi_c)^2 +  \left(u_c K_c + \frac{u_s}{K_s} \right) (\partial_x
    \phi_s)^2 \right] \nonumber \\
&& + \int dx (u_s K_s \Pi_s \partial_x \phi_c -u_c K_c \Pi_c
\partial_x \phi_s)   \nonumber \\
&& + \frac{\Omega}{2\pi a} \int dx \cos \sqrt{2} \theta_s \cos
\sqrt{2} \phi_s  \nonumber \\
&& - \frac{2 g_{1\perp}}{(2\pi a)^2} \int dx \cos \sqrt{8} \phi_s
\end{eqnarray}
Neglecting the spin-charge interaction from the
third line of the Hamiltonian, one finds the Hamiltonian of an XXZ
spin chain in a uniform transverse
field.\cite{giamarchi_spin_flop,nersesyan_2ch} Using a rotation (see
App.~\ref{app:nonabelian}) one can find the ground state of that
Hamiltonian\cite{giamarchi_spin_flop,nersesyan_2ch} and obtain its
correlation functions. In the gapless phase, one finds:
\begin{eqnarray}
\nonumber
&&\langle \rho(j) \rho (j') \rangle  \sim \langle \sigma^z(j) \sigma^z (j')
\rangle  \sim  \frac{(-1)^{j-j'}}{|j-j'|} \\
&+&\frac {1} {2 \pi^2(j-j')^2} \cos \left(\frac{h_s (j-j')}{u_s}
\pm \lambda (j-j') \right) \\
\nonumber
&&\langle  j_\perp(j)  j_\perp(j')\rangle \sim   \frac{(-1)^{j-j'}}{|j-j'|}\\
&&+\frac 1 {2 \pi^2  (j-j')^2}
 \cos \left(\frac{h_s (j-j')}{u_s}  \pm \lambda (j-j')\right),
\end{eqnarray}
with $h_s=O(\Omega)$.
The correlation functions will therefore present
periodic oscillations of wavevector formed of linear combinations of
$\lambda$ and $h_s/u_s$ with integer coefficients, \textit{i. e.}
besides the incommensuration resulting from the flux, a second
incommensuration resulting from interchain hopping is obtained.
At large $\Omega$ a gapped phase can form in which  either the spin-spin or the
conversion current correlation will show a quasi long range order. In such
case, the oscillations associated with the second incommensuration
become exponentially damped, but give rise to Lorentzian-like peaks in
the structure factors.
When $\Omega$ is low, a charge density wave can be
stabilized. Such situation is possible in the case of attractive
interaction, and making attraction between opposite spins stronger is
detrimental to the observation of the second incommensuration. This
explains why, in Ref.~\onlinecite{orignac2017}, we were not
observing a competition of Ising and second incommensuration in the attractive case. At odds, in the repulsive case, the second incommensuration is very robust.

\section{The Hard-core limit}
\label{sec:hard-core}

In this section we report numerical results on the effect of the
interaction between opposite spins when the repulsion between bosons of the same spin is infinite (hard-core case).
Here we focus on the repulsive case, since results obtained in the attractive case have been
discussed in Ref.~\onlinecite{orignac2017}, where we found that instead of having a single
flux-driven Meissner to Vortex transition, the commensurate Meissner phase and the incommensurate Vortex
phase leave space to a Meissner charge-density wave and to a melted vortex phase with short range order.
The transition from the Meissner to the charge density wave phase was
in the Ising universality class, as predicted by fermionization. With a repulsive interaction
we find that the observation of the Ising transition becomes difficult even though signatures of
a vortex melting remain visible.

We show results from DMRG simulations for the filling $\rho = 0.5$ per rung.
We fix interchain hopping  $\Omega/t$ and consider different values of the applied flux $\lambda$ with varying the
 interaction strength $U_\perp$. Simulations are performed in Periodic Boundary Conditions (PBC)
for $L = 32$ and up to $L=64$ in some selected cases, keeping up to $M = 841$ states during
the renormalization procedure. The truncation error, that is the weight of the discarded states,
is at most of order $10^{-5}$, while the error on the ground-state energy is of order $10^{-4}$
at most.

At variance with attractive case, for filling different from unity, in the absence of an applied field
we do not expect the transition from the superfluid Meissner phase to
the density wave phase\cite{hu_pra_2009} since repulsion only gives
rise to a marginally irrelevant perturbation. Thus, the phase diagram
in the presence of flux is expected to be qualitatively different from
the one with attraction.

\begin{figure}[ht]
\centering
\includegraphics[width=9cm]{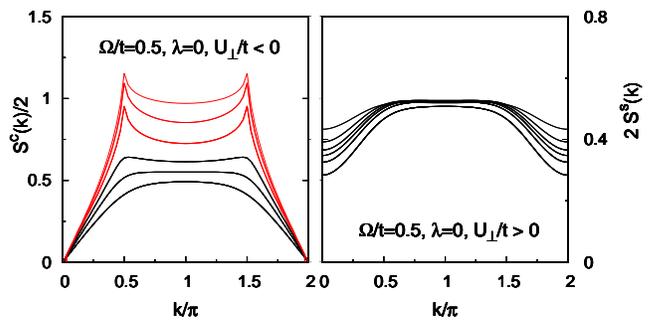}
\caption{Left panel: $S^c(k)$ for the attractive case. Right panel: $S^s(k)$ for the
repulsive case. Interaction strength is $|U_{\perp}|/t=\pm 1.0,1.5,
2.0,3.0 $ and $6.0$ from bottom to top curves. Solid black curves
indicate a Meissner state, while
red solid curves indicate a CDW where the peaks at $k=2 k_F$ develop.
Data from $L=64$ DMRG simulations in PBC at $\lambda=0$ for $\rho=0.5$, at $\Omega/t=0.5$.
Color online.}
\label{fig:notran}
\end{figure}

In Fig.~\ref{fig:notran} we show the response functions, $S^c(k)$ for the attractive case and $S^s(k)$
for the repulsive case, as they evolve upon increasing the strength of interchain interaction,
when $\lambda=0$. As already discussed in Sec.~\ref{sec:bosonization},in the hard-core case, attractive
interchain interaction is expected to give rise to charge density wave, while in the repulsive case
there is not a spin density wave.
In the left panel of Fig.~\ref{fig:notran}, peaks in $S^c(k)$
($U_\perp/t< 0$) develop at $k=\pi/2$ and $k=3\pi/2$  as attraction
increases and the system enters the in-phase density wave phase.
Meanwhile in the right panel of Fig.~\ref{fig:notran},  $S^s(k)$ ($U_\perp/t >0$)
never develops peaks and in fact becomes almost flat as the bosons
become more localized, as repulsion is increased.  Hence, as expected
from marginal irrelevance of interchain repulsion, the spin density wave phase is unfavored.

In order to detect the density wave phases
we choose a value of $\Omega/t$ sufficiently large and an applied flux close to the value at which
the commensurate-incommensurate transition between the Meissner and the Vortex phases occurs in the absence of
interchain interaction.
Let us note that the Luttinger parameters $K_c$ and $K_s$ have a different dependence on
the interchain interaction. In the attractive case $K_c$ is enhanced and $K_s$ is reduced,
thus the region of stability of the Meissner phase is reduced and the system is more prone to reach
the in-phase density wave and vortex regime. On the contrary, in the repulsive case $K_c$ is reduced
and $K_s$ is increased and, as a consequence, the Meissner phase becomes more stable at the expense of the Vortex
and density wave ones.

\begin{figure}[ht]
\centering
\includegraphics[width=9cm]{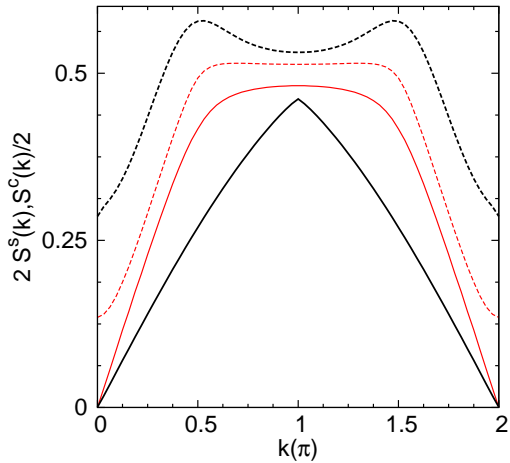}
\caption{ $S^c(k)$and $S^s(k)$ for the repulsive case, respectively solid and dotted lines, for
$U_\perp/t=0.5$ and $5.5$, red and black curves respectively. Data from $L=64$ DMRG simulations in PBC
at $\lambda=0.0625 \pi$, for $\Omega/t=0.125$ and $\rho=0.5$. Color online.}
\label{fig:sk_before}
\end{figure}

We consider the following case: $\Omega/t=0.125$ at
different applied fluxes. At $\lambda \lesssim \lambda_c(U_\perp=0)$, {\it.i.e} just before the C-IC transition
occurs, the system never develops a density wave.
In Fig.~\ref{fig:sk_before} we show the behavior of the spin and the charge response functions $S^s(k)$
and $S^c(k)$ respectively, for small and large interaction interaction strength.
On increasing the strength the spin static structure factor develops shoulders at $k=2 k_F=\pm \pi/2$ signaling
the incipient transition towards a density wave phase, while the static structure factor for low
momentum show the expected linear behavior $S^c(k)\simeq \frac{2 K_c}{\pi}|k|$ for gapless
charge excitations.
$K_c$ smoothly decreases as a function of $U_\perp$, going from one, as for a
non-interacting hard-core Bose system, towards $1/2$ as shown from the slope of the low momentum
linear behavior of $S^c(k)$ (see Fig.~\ref{fig:sk_before}).

\begin{figure}[ht]
\centering
\includegraphics[width=9cm]{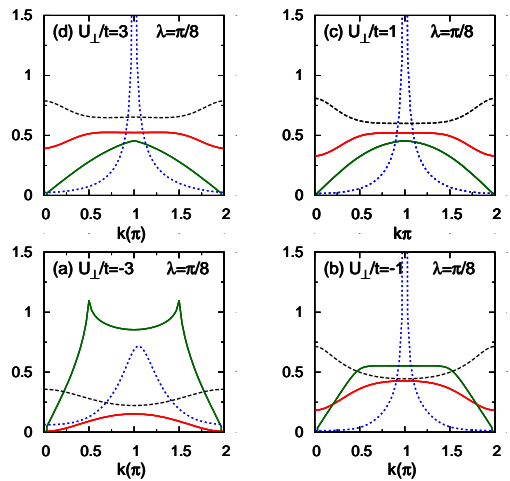}
\caption{ $S^c(k)$ and $S^s(k)$ are respectively shown as dark-green and red solid lines in all panels.
Blue dotted lines are for spin resolved momentum distribution $n_\sigma(k)$ whose argument has been
shifted shifted of $\pi$ and black dashed lines are for rung-rung correlation function $C(k)$.
Panels $a,b,c$ and $d$ are respectively for $U_\perp=-3,-1,1$ and $3$. Data from $L=64$ DMRG simulations in PBC
at $\lambda=\pi/8$, for $\Omega/t=0.5$ and $\rho=0.5$. Color online.}
\label{fig:small_lambda}
\end{figure}

The asymmetry between the attractive and the repulsive case persists
in the presence of an applied flux, as shown in
Fig.~\ref{fig:small_lambda}, where we follow the response functions
change when we increase the interaction strength at fixed $\lambda$ and $\Omega/t=0.5$.
At small $|U_\perp|$, panel $(c)$ and $(b)$ we start from Meissner phase where the momentum distribution has a single peak at $k=0$, but for larger interaction strength, while in the attractive case we are in a melted Vortex phase, panel $(a)$, in the repulsive case the system
is still in the Meissner phase and $S^s(k)$ shows only shoulders at $k=2 k_F$ (panel $(d)$).

In the following we investigate the system for fixed $\Omega/t=0.125$ and at a fixed applied flux
for which the system is in the Vortex state in the absence of interaction between the chains ($U_\perp=0$).
\begin{figure}[ht]
\centering
\includegraphics[width=9cm]{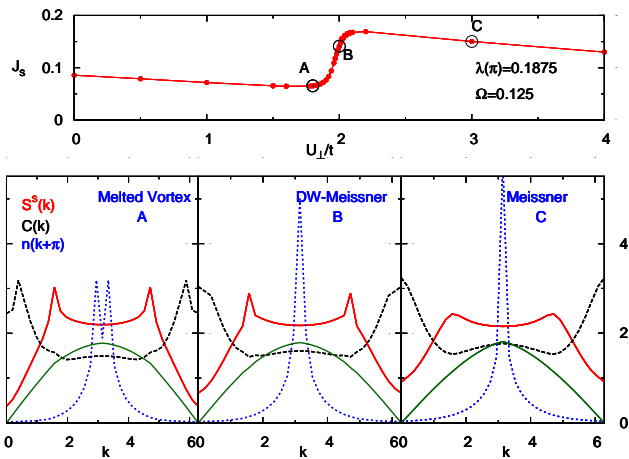}
\caption{ Upper panel shows the spin current $J_s$ as a function of strength of interchain interaction,
solid red line is only a guide to eye. Panels below show $ 8 S^s(k)$ (red solid line), $ 4 C(k)$
(black dashed line), and $n(k)$ (blue dotted line) where the argument of this
last quantity has been shifted of $\pi$. Left, center and right panel shows these quantities for the
cases indicated by the point A, B and C in the upper panel, respectively corresponding to cases where
the system is in the melted Vortex phase, in the CDW--Meissner phase and in the Meissner phase.
Data shown are from DMRG simulations in PBC for L=32. Color online.}
\label{fig:across_o0.125}
\end{figure}
In the absence of the interaction the spin response function $S^s(k)$ displays the expected linear
behavior at small momentum and a discontinuity in the derivative at $k=2 k_F$ \cite{our_2016}.
As we increase the interaction strength (panel A in Fig.~\ref{fig:across_o0.125}) the spin structure
factor develops peaks at $k=\pi/2$ and $k=3\pi/2$ and an almost quadratic behavior at small wavevector.
The quadratic behavior indicates that spin excitations remain gapped, while the
presence of peaks at $k=\pi/2,3\pi/2$ is the signature of a zig-zag
charge density wave (in the ladder language) or a  spin density wave (in the spin-orbit language).
The momentum distribution as well the rung-rung response
function $C(k)$ develop two separate peaks indicating the presence of
an incommensuration. Thus, we can identify the phase to the so-called
melted Vortex phase.\cite{orignac2017}
For large value of interaction, panel $C$ of Fig.~\ref{fig:across_o0.125}, the system is in
strongly correlated Meissner phase, indeed momentum distribution show only one peak centered
around $k=0$, and $S^s(k)$ shows the incipient transition towards
the CDW-Meissner phase in its spin response.
In panel B, we have an intermediate situation, where the DW peaks are still visible in the
spin response function, but not incommensuration.
We conjecture that this corresponds to the so-called charge-density Meissner phase.

In the upper panel of Fig.~\ref{fig:across_o0.125} we show the spin current $J_s$ as a
function of the strength of interchain interaction when the system goes from the Vortex state to the Meissner one:
there is no cusp indicating a square root threshold singularity typical of the C-IC
transition, instead the spin current only shows at most a
vertical tangent indicating a possible logarithmic divergence of its derivative.
To summarize, under application of interleg repulsion, the Vortex
phase becomes first a melted vortex phase via a BKT transition, then
past the disorder point a DW-Meissner is formed, and finally the Meissner state is stabilized at large repulsion.


As discussed in the previous section, in the presence of the so-called
second incommensuration,\cite{our_2015,our_2016}
the picture becomes more complicated. Indeed, in such a case, nearby $\lambda \simeq \pi n$ there is a new
incommensurate wavevector which gives, in the various structure factors, extra peaks whose magnitude of
which is controlled by $\Omega$.
In order to illustrate that situation we have made simulations for a larger interchain hopping, namely
$\Omega/t=0.5$, so that we have the C-CI transition nearby  $\lambda= n \pi$ in absence of
interchain interaction and therefore near the occurrence of the second incommensuration.

\begin{figure}[ht]
\centering
\includegraphics[width=9cm]{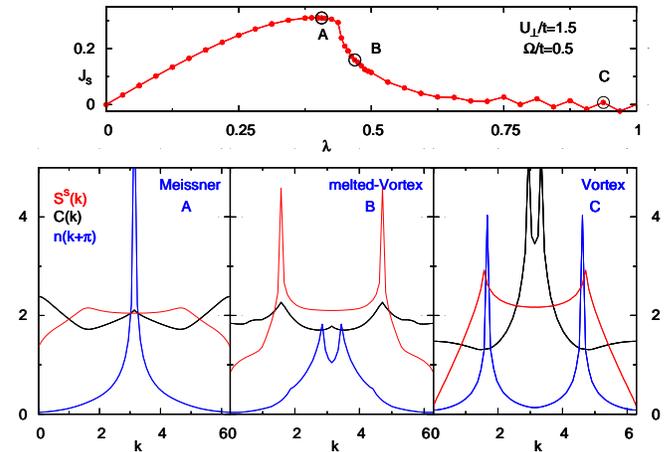}
\caption{Upper panel shows the spin current $J_s$ as a function of the applied flux, red line is only a guide
to eye. Panels below show $S^s(k)$ (red solid line), $C(k)$ (black line), $n_\sigma(k)$ (blue line)
where the argument of this last quantity has been shifted of $\pi$. Left, center and right panels show these
quantities for the cases indicated by the point $A$, $B$ and $C$ in the upper panel, respectively corresponding
to cases where the system is in the Meissner phase, in the melted-Vortex phase and in the Vortex phase.
Data shown are from DMRG simulations in PBC for L=64. Color online.}
 \label{fig:across_u1.5}
\end{figure}

In Fig.~\ref{fig:across_u1.5} situation at fixed $U_\perp/t=1.5$ and $\Omega/t=0.5$ is shown. In the upper
panel we follow the spin current as a function of the applied field. It shows the typical behavior
of the Meissner phase when it increases as a function of $\lambda$, then it rapidly decreases when entering
the Vortex phase which is however short-ranged ordered
and finally for larger $\lambda$ enter the quasi long range ordered
Vortex phase, as it can also seen from the typical finite size induced oscillations
in this quantity.\cite{didio2015a}
The Meissner phase is shown in panel $A$, while the melted-Vortex phase is shown in panel $B$, where
the spin response function has the expected peaks at $k=\pi/2$ and $3 \pi/2$, yet it has the low momentum
behavior observed in the presence of a second incommensuration.\cite{our_2016} In this case,
in the momentum distribution is possible to see besides the primary peaks also the secondary peaks related to the second incommensuration.
These peaks can be seen also in the rung-rung correlation function $C(k)$. However, both of these functions
do not show appreciable size effects attesting the short range of the incommensurate order.
In panel $C$ we recover the quasi-long range ordered Vortex phase.

As a last comment we want to stress the fact that in the rung-rung current correlation function in the
Meissner phase, see panel $C$ of Fig.~\ref{fig:across_o0.125} and panel $A$ of Fig.~\ref{fig:across_u1.5}, shows
respectively a Lorentzian-like peak and a cusp centered at $k=4k_F=\pi$ as the result of higher order term
in the Haldane expansion when we derive the rung current. This cusp is present since
the exponent $K_c$ is decreasing with repulsion, thus enhancing the
contribution of the contribution at $\pi$ compared with the attractive
case.

\section{Conclusions}
\label{sec:ccl}

To conclude, we have analyzed the phase diagram of boson ladder in the presence of an artificial gauge field, when a repulsive interchain interaction is switched on.
We have shown, using bosonization, fermionization and DMRG approach, that the
the commensurate-incommensurate transition between the Meissner phase
and the QLRO vortex phase is replaced by an Ising-like transition
towards a commensurate zig-zag density wave phase.
The fermionization approach has allowed us
to predict the existence of a disorder point after which
the bosonic Green's functions and the rung current correlation function
develop exponentially damped oscillations in real space while zig-zag density wave phase persists. This phase is recognized as a melted vortex phase. Differently from
the attractive interaction, a second-incommensuration, i.e. an extra periodic oscillation of the correlation functions at wavevectors formed
by a linear combinations of the flux and the interchain interaction,
dominates even away from half-filling. As numerically shown, the hard
core limit in the chains favors the zig-zag density wave phase.
Our predictions on the melting of vortices in Bose-Einstein
condensates and on the second incommensuration in optical lattices can be traced in current experiments by the measuring the static structure factors and momentum
distributions, together with the rung current.

\section{Acknowledgments}
We acknowledge A. Celi, M. Calvanese Strinati and E. Tirrito for fruitful discussions.
Simulations were performed at Universit\`a di Salerno, Universit\`a di Trieste and Democritos local computing facilities. M. Di Dio and S. De Palo thank F. Ortolani for the DMRG code.
E. Orignac acknowledges hospitality from Universit\`a of Salerno.
\appendix

\section{Hard core boson limit and mappings}
\label{sec:hcb-case}
That limit corresponds to
$U_{\uparrow\uparrow}=U_{\downarrow\downarrow} \to +\infty$. In that
limit, the bosonic ladder can be mapped to an anisotropic  two-leg
ladder model with Dzyaloshinskii-Moriya\cite{dzyaloshinskii_interaction,moryia_asym_int} interaction, and to the
Hubbard model.

\subsection{Mapping to a spin ladder}
\label{sec:ladder-mapping}

If we consider hard core bosons, we can use the
mapping of hard core bosons to spins 1/2:
\begin{eqnarray}
  \label{eq:boson-to-spin}
  b_j^\dagger = S_j^+ \\
 b_j = S_j^{-} \\
 b^\dagger_j b_j = S_j^z+\frac 1 2,
\end{eqnarray}
 which can be deduced easily from the Holstein-Primakoff
 representation\cite{holstein40_operators} of  spin-1/2 operators.
With such mapping, we can rewrite the
Hamiltonian~(\ref{eq:full-lattice-ham}) as a two-leg ladder
Hamiltonian in which the upper and the lower leg have uniform
Dzyaloshinskii Moriya interaction. In the two leg ladder
representation, $\Omega$ and $U_{\uparrow \downarrow}$ become the rung
exchange interaction, $t\cos (\lambda/2) $ and
$U_{\uparrow\uparrow},U_{\uparrow\uparrow}$ become the leg exchange
interaction, $t\sin (\lambda/2) $ becomes the Dzyaloshinskii-Moriya term.

\subsection{Mapping to spin-1/2 fermions}
\label{sec:hubbard}

Another possible mapping in the case of hard core bosons $U_{\uparrow
  \uparrow},  U_{\downarrow
  \downarrow} \to \infty$ is to the Hubbard model. This mapping is only
valid when $\Omega=0$, but it allows to take advantage of the
integrability of the Hubbard
model.\cite{lieb_hubbard_exact,frahm_confinv,frahm_confinv_field,andrei_trieste93}
The mapping, is obtained from the Jordan-Wigner
transformation\cite{jordan_transformation} that maps hard core bosons
operators $b_{j\sigma}$
to fermion operators $c_{j\sigma}$ :
\begin{eqnarray}
  \label{eq:jordan-wigner}
  b_{j\sigma}&=&\eta_\sigma c_{j,\sigma} e^{i \pi \sum_{k<j} c^\dagger_{k,\sigma}
    c_{k,\sigma}},  \\
   b^\dagger_{j\sigma}  b_{j\sigma}&=&c^\dagger_{j,\sigma}
   c_{j,\sigma},
\end{eqnarray}
where $\{\eta_\sigma,\eta_{\sigma'}\}_+=\delta_{\sigma\sigma'}$.
The Hamiltonian~(\ref{eq:full-lattice-ham}) with
$\Omega=0$ is rewritten as:

\begin{eqnarray}
  \label{eq:hubbard-form}
  H=-t \sum_{j,\sigma} (c^\dagger_{j+1,\sigma} e^{-i\lambda \sigma}
  c_{j,\sigma} + \mathrm{H. c.})    + U \sum_j n_{j,\uparrow} n_{j,\downarrow}
\end{eqnarray}

The gauge transformation\cite{zvyagin2012}  $c_{j,\sigma} = e^{-i \lambda
  \sigma j} a_{j,\sigma}$ reduces the
Hamiltonian~(\ref{eq:hubbard-form}) to the Hubbard
form. The Hubbard model presents a spin-charge separation. When
interactions are repulsive, and away from half-filling, charge and
spin modes are gapless, whereas with attractive interactions charge
modes are always gapless but spin modes are gapped.
In terms of the original bosons, total density modes are always
gapless away from half-filling, but the chain antisymmetric density
fluctuations are gapped with attractive interaction giving rise to a
symmetric density wave phase, gapless with
repulsive interaction.

\section{Asymptotic behavior of the Green's  functions}
\label{app:gf}
To estimate the asymptotic behavior of the Green's functions, we
apply a contour integral method\cite{bender78_book} to the integral
\begin{eqnarray}\label{eq:i1-bar}
  \bar{I}_1(x)= \int_{-\infty}^{\infty} \frac{dk}{2\pi} \frac{e^{ik |x|}}{E_-(k)}.
\end{eqnarray}
The function $E_-(k)$ has only branch cut singularities in the upper
half plane. The branch cuts arise either from
 $h^2 (uk)^2 + h^2 m^2 + m^2 \Delta^2 <0$ or
 $(uk)^2 + m^2 + \Delta^2 + h^2 -2 \sqrt{h^2 (uk)^2 + h^2 m^2 +
    m^2 \Delta^2} <0$.
The first branch cut, obtained for $u^2 k^2 < -m^2 (1+\Delta^2/h^2)$ gives
a contribution decaying as $e^{-m \sqrt{1+\Delta^2/h^2}|x|/u}$, that
can be ignored for $|x| \gg m/u$.
The contribution of the cuts of the second type depends whether $h<m$
or $h>m$.
For $h<m$, there is a single branch cut extending along the imaginary
axis from $i|\Delta -\sqrt{m^2-h^2}|/u< k < i(\Delta
+\sqrt{m^2-h^2})$. We can rewrite the integral (\ref{eq:i1-bar}) as:
\begin{eqnarray}
\bar{I}_1(x) = \int_{\frac{|\Delta-\sqrt{m^2-h^2}|} u}^{\frac{\Delta+\sqrt{m^2-h^2}} u} \frac{dk}{\pi} \frac{e^{-k |x|}}{E_-(ik)}.
\end{eqnarray}
showing that $I_1(x) \sim e^{-\frac{|\Delta-\sqrt{m^2-h^2}||x|}
  u}$. This gives a  correlation length diverging as $\sim
|m-\sqrt{h^2+\Delta^2}|^{-1}$ near the Ising transition.

\begin{figure}[h]
  \centering
  \includegraphics[width=9cm]{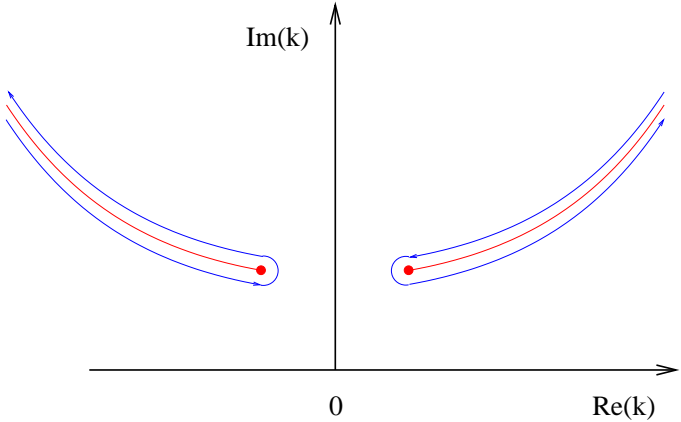}
  \caption{The integration path in complex $k$ plane for $h>m$. The red curves are such that $E_-(k)^2<0$. The red dots at extremities of the curve are the points where $E_-(k)=0$. }
  \label{fig:contour-i1bar}
\end{figure}

For $h>m$, there are two branch cuts given by:
\begin{eqnarray}
  \sqrt{(uk)^2 + m^2 + \frac{m^2 \Delta^2}{h^2}} = h \pm i \Delta
  \sqrt{1-\frac{m^2}{h^2}} \cosh \alpha,
\end{eqnarray}
and $\alpha$ real. The integration path in the complex plane is
represented on Fig.~\ref{fig:contour-i1bar}. The branch cuts terminate
at the branch points $k_d^{(\pm)}=i \frac \Delta u \pm \frac{\sqrt{h^2-m^2}}
u$ such that $E_-(k_d^{(\pm)})^2=0$. The long distance behavior of
$\bar{I}_1$ is determined by these two branch points as:
\begin{equation}
  \label{eq:disorder-i1bar}
  \bar{I}_1(x) \sim e^{-\frac{\Delta |x|} u} \left[e^{i
      \frac{\sqrt{h^2-m^2}|x|} u} \varphi_1(x) + e^{-i
      \frac{\sqrt{h^2-m^2}|x|} u} \varphi_1(x)^*\right],
\end{equation}
so that oscillations of wavevector $\sqrt{h^2-m^2}/u$ appear in the
real space correlation functions for $h>m$. The point
$h=m$ is called a disorder
point\cite{stephenson1970a,stephenson1970b}.
Disorder points are  known to occur in frustrated quantum Ising chains in
transverse field,\cite{beccaria2006}
bilinear-biquadratic spin-1
chains,\cite{golinelli_incommensurate,schollwoeck1996}  frustrated
spin-1/2 \cite{bursill1995,deschner2013} and spin-1
\cite{pixley2014,chepiga2016} chains. 
They can be classified\cite{stephenson1970b} into disorder points of the first kind (with
parameter dependent incommensuration) and
disorder point of the second kind (with parameter independent
incommensuration). In our model, the disorder point is of the first
kind.

\section{Second incommensuration and canonical
  transformation}\label{app:nonabelian}
In this section, we give some details on the
rotation\cite{giamarchi_spin_flop,nersesyan_2ch} used to
diagonalize the Hamiltonian obtained after the unitary
transformation of Eq.~(\ref{eq:unitary-op}). First, we rewrite our
Hamiltonian~(\ref{eq:u-trasf-h})  using nonabelian
bosonization:\cite{witten_wz}

\begin{eqnarray}
&& H_s=\frac{2\pi v} 3 \int dx (\vec{J}_R \cdot \vec{J}_R +\vec{J}_L
 \cdot \vec{J}_L) + g_{1\parallel} \int dx  J_R^z J_L^z
\\
&& + g_{1\perp}  \int dx   (J_R^x
  J_L^x+J_R^y J_L^y) + \Omega \int dx  (J_R^y+J_L^y)
\end{eqnarray}
with $g_{1\parallel}\ne g_{1\perp}$ .
Using a $\frac \pi 2$ rotation around the $y$ axis\cite{our_2015} we can rewrite:
\begin{eqnarray}
&&  H_s=\frac{2\pi v} 3 \int dx (\vec{\tilde{J}}_R \cdot \vec{\tilde{J}}_R +\vec{\tilde{J}}_L
  \cdot \vec{\tilde{J}}_L) + g_{1\parallel} \int dx  \tilde{J}_R^y\tilde{J}_L^y \nonumber \\
&&  + g_{1\perp}   \int dx  (\tilde{J}_R^x
  \tilde{J}_L^x+\tilde{J}_R^z \tilde{J}_L^z) + \Omega \int dx  (\tilde{J}_R^z+\tilde{J}_L^z)
\end{eqnarray}
Finally, returning to abelian bosonization, we
obtain\cite{giamarchi_spin_flop,nersesyan_2ch}
\begin{eqnarray}
 H_s&=& \int \frac{dx}{2\pi}\left[u K (\pi \tilde{\Pi}_s)^2 + \frac{u}{K}
    (\partial_x \tilde{\phi}_s)^2\right] - \frac{\Omega}{\pi \sqrt{2}} \int dx
  \partial_x \tilde{\phi}_s \nonumber \\
&&+ \frac{2(g_{1\perp}+g_{1\parallel})}{(2\pi a)^2} \int dx \cos
\sqrt{8} \tilde{\phi}_s \nonumber \\
&&+ \frac{2(g_{1\perp}-g_{1\parallel})}{(2\pi a)^2} \int dx \cos
\sqrt{8} \tilde{\theta}_s
\end{eqnarray}
As we can see, either we obtain a fixed point with
$\tilde{\theta}_s$ long range ordered or a gapless fixed point.
In both cases since we may eliminate the $\partial_x \tilde{\phi}_s$ by a shift of the $\tilde{\phi}_s$ field, one has $\langle \tilde{\phi_s} \rangle =
\frac{h_s}{\sqrt{2} u_s} x$. When $\tilde{\theta}_s$ is gapless, this gives
rise to the second incommensuration of Ref.~\onlinecite{our_2016}.
To be more precise, if we consider the bosonized expression for the observables:
\begin{eqnarray}
  U^\dagger \rho(x) U &=& \rho_0 -\frac{\sqrt{2}}\pi \partial_x \phi_c + \cos
  \sqrt{2} (\phi_c - 2 \pi \rho_0 x) \cos \sqrt{2} \phi_s \\
   U^\dagger \sigma^z(x) U& =&  -\frac{\sqrt{2}}\pi \partial_x \phi_s + \cos
  \sqrt{2} (\phi_c - 2 \pi \rho_0 x) \sin \sqrt{2} \phi_s \\
U^\dagger j_\perp(x) U &=& \frac{\Omega}{\pi a} \left[ \sum_{r=\pm 1}
   \sin \sqrt{2} (\theta_s + r \phi_s) \cos \sqrt{2} \phi_c
  \right. \nonumber \\  \nonumber
&+& \cos\sqrt{2} (\theta_s + r \phi_s) \sin \sqrt{2} \phi_c
 + \sin \sqrt{2}   \theta_s \cos (\sqrt{2} \phi_c +\lambda x) \\
\nonumber
&+& \left. \cos \sqrt{2} \theta_s \sin(
   \sqrt{2} \phi_c +\lambda x) \right]
\end{eqnarray}
and perform here the shift of the field $\tilde{\phi}_s\rightarrow \phi_s- \frac{h_s x}{\sqrt{2} u_s} x$
and using a rotation of the $SU(2)_1$ primary fields\cite{our_2015}, we reexpress the observables as:
\begin{eqnarray}
 U^\dagger \rho(x) U &=& \rho_0 -\frac{\sqrt{2}}\pi \partial_x \phi_c
 \nonumber \\
&&+ \cos\sqrt{2} (\phi_c - 2 \pi \rho_0 x) \cos \left(\sqrt{2}
    \tilde{\phi_s} + \frac{h_s}{u_s} x\right)  \\
  U^\dagger \sigma^z(x) U &=&  -\frac 1 {\pi a}
  \sum_{r,r'=\pm} e^{i r\sqrt{2} (\tilde{\theta}_s+  r'\tilde{\phi}_s) + i rr'
      \frac{h_s x}{u_s}}  \nonumber\\
&&- \cos
  \sqrt{2} (\phi_c - 2 \pi \rho_0 x) \cos \sqrt{2} \tilde{\theta}_s  \\
U^\dagger j_\perp(x) U &=&  \frac{\Omega}{\pi a} \left[ \sum_{r=\pm 1}
   \sin \sqrt{2} \left(\tilde{\theta}_s + r \tilde{\phi}_s
   +\frac{h_s}{u_s} x \right) \cos \sqrt{2} \phi_c \right.\nonumber \\
-\frac{1}{\pi\sqrt{2}} &&\partial_x \tilde{\phi_s}  \sin \sqrt{2} \phi_c
+ \sin \sqrt{2} \tilde{\theta}_s \cos (\sqrt{2} \phi_c +\lambda x) \nonumber\\
&&+ \left. \sin
   \left(\sqrt{2} \tilde{\phi}_s +\frac{h_s}{u_s} x\right) \sin(
   \sqrt{2} \phi_c +\lambda x) \right]
\end{eqnarray}

In the gapless case, taking the expectation value gives the second
incommensuration.


\begin{thebibliography}{118}
\expandafter\ifx\csname natexlab\endcsname\relax\def\natexlab#1{#1}\fi
\expandafter\ifx\csname bibnamefont\endcsname\relax
  \def\bibnamefont#1{#1}\fi
\expandafter\ifx\csname bibfnamefont\endcsname\relax
  \def\bibfnamefont#1{#1}\fi
\expandafter\ifx\csname citenamefont\endcsname\relax
  \def\citenamefont#1{#1}\fi
\expandafter\ifx\csname url\endcsname\relax
  \def\url#1{\texttt{#1}}\fi
\expandafter\ifx\csname urlprefix\endcsname\relax\def\urlprefix{URL }\fi
\providecommand{\bibinfo}[2]{#2}
\providecommand{\eprint}[2][]{\url{#2}}

\bibitem[{\citenamefont{Jaksch and Zoller}(2005)}]{jaksch05_coldatoms}
\bibinfo{author}{\bibfnamefont{D.}~\bibnamefont{Jaksch}} \bibnamefont{and}
  \bibinfo{author}{\bibfnamefont{P.}~\bibnamefont{Zoller}},
  \bibinfo{journal}{Ann. Phys. (N. Y.)} \textbf{\bibinfo{volume}{315}},
  \bibinfo{pages}{52} (\bibinfo{year}{2005}), \bibinfo{note}{cond-mat/0410614}.

\bibitem[{\citenamefont{Lewenstein et~al.}(2007)\citenamefont{Lewenstein,
  Sanpera, Ahufinger, Damski, {Sen De}, and
  Sen}}]{lewenstein07_coldatoms_review}
\bibinfo{author}{\bibfnamefont{M.}~\bibnamefont{Lewenstein}},
  \bibinfo{author}{\bibfnamefont{A.}~\bibnamefont{Sanpera}},
  \bibinfo{author}{\bibfnamefont{V.}~\bibnamefont{Ahufinger}},
  \bibinfo{author}{\bibfnamefont{B.}~\bibnamefont{Damski}},
  \bibinfo{author}{\bibfnamefont{A.}~\bibnamefont{{Sen De}}}, \bibnamefont{and}
  \bibinfo{author}{\bibfnamefont{U.}~\bibnamefont{Sen}}, \bibinfo{journal}{Ann.
  Phys. (N. Y.)} \textbf{\bibinfo{volume}{56}}, \bibinfo{pages}{243}
  (\bibinfo{year}{2007}), \bibinfo{note}{cond-mat/0606771}.

\bibitem[{\citenamefont{Bloch et~al.}(2008)\citenamefont{Bloch, Dalibard, and
  Zwerger}}]{bloch08_manybody}
\bibinfo{author}{\bibfnamefont{I.}~\bibnamefont{Bloch}},
  \bibinfo{author}{\bibfnamefont{J.}~\bibnamefont{Dalibard}}, \bibnamefont{and}
  \bibinfo{author}{\bibfnamefont{W.}~\bibnamefont{Zwerger}},
  \bibinfo{journal}{Rev. Mod. Phys.} \textbf{\bibinfo{volume}{80}},
  \bibinfo{pages}{885} (\bibinfo{year}{2008}).

\bibitem[{\citenamefont{{Cazalilla} et~al.}(2011)\citenamefont{{Cazalilla},
  {Citro}, {Giamarchi}, {Orignac}, and {Rigol}}}]{cazalilla2011}
\bibinfo{author}{\bibfnamefont{M.~A.} \bibnamefont{{Cazalilla}}},
  \bibinfo{author}{\bibfnamefont{R.}~\bibnamefont{{Citro}}},
  \bibinfo{author}{\bibfnamefont{T.}~\bibnamefont{{Giamarchi}}},
  \bibinfo{author}{\bibfnamefont{E.}~\bibnamefont{{Orignac}}},
  \bibnamefont{and} \bibinfo{author}{\bibfnamefont{M.}~\bibnamefont{{Rigol}}},
  \bibinfo{journal}{Rev. Mod. Phys.} \textbf{\bibinfo{volume}{83}},
  \bibinfo{pages}{1405} (\bibinfo{year}{2011}).

\bibitem[{\citenamefont{Lin et~al.}(2011)\citenamefont{Lin, Jimenez-Garcia, and
  Spielman}}]{lin2011_soc}
\bibinfo{author}{\bibfnamefont{Y.}~\bibnamefont{Lin}},
  \bibinfo{author}{\bibfnamefont{K.}~\bibnamefont{Jimenez-Garcia}},
  \bibnamefont{and} \bibinfo{author}{\bibfnamefont{I.~B.}
  \bibnamefont{Spielman}}, \bibinfo{journal}{Nature (London)}
  \textbf{\bibinfo{volume}{471}}, \bibinfo{pages}{83} (\bibinfo{year}{2011}).

\bibitem[{\citenamefont{Dalibard et~al.}(2011)\citenamefont{Dalibard, Gerbier,
  Juzeli{\=u}nas, and {\"O}hberg}}]{dalibard2011gauge}
\bibinfo{author}{\bibfnamefont{J.}~\bibnamefont{Dalibard}},
  \bibinfo{author}{\bibfnamefont{F.}~\bibnamefont{Gerbier}},
  \bibinfo{author}{\bibfnamefont{G.}~\bibnamefont{Juzeli{\=u}nas}},
  \bibnamefont{and}
  \bibinfo{author}{\bibfnamefont{P.}~\bibnamefont{{\"O}hberg}},
  \bibinfo{journal}{Rev. Mod. Phys.} \textbf{\bibinfo{volume}{83}},
  \bibinfo{pages}{1523} (\bibinfo{year}{2011}).

\bibitem[{\citenamefont{{Galitski} and {Spielman}}(2013)}]{galitski2013_soc}
\bibinfo{author}{\bibfnamefont{V.}~\bibnamefont{{Galitski}}} \bibnamefont{and}
  \bibinfo{author}{\bibfnamefont{I.~B.} \bibnamefont{{Spielman}}},
  \bibinfo{journal}{Nature (London)} \textbf{\bibinfo{volume}{494}},
  \bibinfo{pages}{49} (\bibinfo{year}{2013}).

\bibitem[{\citenamefont{Celi et~al.}(2014)\citenamefont{Celi, Massignan,
  Ruseckas, Goldman, Spielman, Juzeli\ifmmode~\bar{u}\else \={u}\fi{}nas, and
  Lewenstein}}]{celi2014}
\bibinfo{author}{\bibfnamefont{A.}~\bibnamefont{Celi}},
  \bibinfo{author}{\bibfnamefont{P.}~\bibnamefont{Massignan}},
  \bibinfo{author}{\bibfnamefont{J.}~\bibnamefont{Ruseckas}},
  \bibinfo{author}{\bibfnamefont{N.}~\bibnamefont{Goldman}},
  \bibinfo{author}{\bibfnamefont{I.~B.} \bibnamefont{Spielman}},
  \bibinfo{author}{\bibfnamefont{G.}~\bibnamefont{Juzeli\ifmmode~\bar{u}\else
  \={u}\fi{}nas}}, \bibnamefont{and}
  \bibinfo{author}{\bibfnamefont{M.}~\bibnamefont{Lewenstein}},
  \bibinfo{journal}{Phys. Rev. Lett.} \textbf{\bibinfo{volume}{112}},
  \bibinfo{pages}{043001} (\bibinfo{year}{2014}).

\bibitem[{\citenamefont{Livi et~al.}(2016)\citenamefont{Livi, Cappellini, Diem,
  Franchi, Clivati, Frittelli, Levi, Calonico, Catani, Inguscio et~al.}}]{Livi}
\bibinfo{author}{\bibfnamefont{L.~F.} \bibnamefont{Livi}},
  \bibinfo{author}{\bibfnamefont{G.}~\bibnamefont{Cappellini}},
  \bibinfo{author}{\bibfnamefont{M.}~\bibnamefont{Diem}},
  \bibinfo{author}{\bibfnamefont{L.}~\bibnamefont{Franchi}},
  \bibinfo{author}{\bibfnamefont{C.}~\bibnamefont{Clivati}},
  \bibinfo{author}{\bibfnamefont{M.}~\bibnamefont{Frittelli}},
  \bibinfo{author}{\bibfnamefont{F.}~\bibnamefont{Levi}},
  \bibinfo{author}{\bibfnamefont{D.}~\bibnamefont{Calonico}},
  \bibinfo{author}{\bibfnamefont{J.}~\bibnamefont{Catani}},
  \bibinfo{author}{\bibfnamefont{M.}~\bibnamefont{Inguscio}},
  \bibnamefont{et~al.}, \bibinfo{journal}{Phys. Rev. Lett.}
  \textbf{\bibinfo{volume}{117}}, \bibinfo{pages}{220401}
  (\bibinfo{year}{2016}).

\bibitem[{\citenamefont{{Regnault} and {Jolicoeur}}(2003)}]{regnault2003_qhe}
\bibinfo{author}{\bibfnamefont{N.}~\bibnamefont{{Regnault}}} \bibnamefont{and}
  \bibinfo{author}{\bibfnamefont{T.}~\bibnamefont{{Jolicoeur}}},
  \bibinfo{journal}{Phys. Rev. Lett.} \textbf{\bibinfo{volume}{91}},
  \bibinfo{pages}{030402} (\bibinfo{year}{2003}),
  \eprint{arXiv:cond-mat/0212477}.

\bibitem[{\citenamefont{Atala et~al.}(2014)\citenamefont{Atala, Aidelsburger,
  Lohse, Barreiro, Paredes, and Bloch}}]{atala2014}
\bibinfo{author}{\bibfnamefont{M.}~\bibnamefont{Atala}},
  \bibinfo{author}{\bibfnamefont{M.}~\bibnamefont{Aidelsburger}},
  \bibinfo{author}{\bibfnamefont{M.}~\bibnamefont{Lohse}},
  \bibinfo{author}{\bibfnamefont{J.}~\bibnamefont{Barreiro}},
  \bibinfo{author}{\bibfnamefont{B.}~\bibnamefont{Paredes}}, \bibnamefont{and}
  \bibinfo{author}{\bibfnamefont{I.}~\bibnamefont{Bloch}},
  \bibinfo{journal}{Nat. Phys.} \textbf{\bibinfo{volume}{10}},
  \bibinfo{pages}{588} (\bibinfo{year}{2014}).

\bibitem[{\citenamefont{Kardar}(1986)}]{kardar_josephson_ladder}
\bibinfo{author}{\bibfnamefont{M.}~\bibnamefont{Kardar}},
  \bibinfo{journal}{Phys. Rev. B} \textbf{\bibinfo{volume}{33}},
  \bibinfo{pages}{3125} (\bibinfo{year}{1986}).

\bibitem[{\citenamefont{Orignac and
  Giamarchi}(2001{\natexlab{a}})}]{orignac01_meissner}
\bibinfo{author}{\bibfnamefont{E.}~\bibnamefont{Orignac}} \bibnamefont{and}
  \bibinfo{author}{\bibfnamefont{T.}~\bibnamefont{Giamarchi}},
  \bibinfo{journal}{Phys. Rev. B} \textbf{\bibinfo{volume}{64}},
  \bibinfo{pages}{144515} (\bibinfo{year}{2001}{\natexlab{a}}).

\bibitem[{\citenamefont{Cha and Shin}(2011)}]{cha2011}
\bibinfo{author}{\bibfnamefont{M.-C.} \bibnamefont{Cha}} \bibnamefont{and}
  \bibinfo{author}{\bibfnamefont{J.-G.} \bibnamefont{Shin}},
  \bibinfo{journal}{Phys. Rev. A} \textbf{\bibinfo{volume}{83}},
  \bibinfo{pages}{055602} (\bibinfo{year}{2011}).

\bibitem[{\citenamefont{Ambegaokar et~al.}(1982)\citenamefont{Ambegaokar,
  Eckern, and {Sch\"on}}}]{ambegaokar82_josephson_dissipation}
\bibinfo{author}{\bibfnamefont{V.}~\bibnamefont{Ambegaokar}},
  \bibinfo{author}{\bibfnamefont{U.}~\bibnamefont{Eckern}}, \bibnamefont{and}
  \bibinfo{author}{\bibfnamefont{G.}~\bibnamefont{{Sch\"on}}},
  \bibinfo{journal}{Phys. Rev. Lett.} \textbf{\bibinfo{volume}{48}},
  \bibinfo{pages}{1745} (\bibinfo{year}{1982}).

\bibitem[{\citenamefont{{Korshunov}}(1989)}]{korshunov_dissipative_josephson1d}
\bibinfo{author}{\bibfnamefont{S.~E.} \bibnamefont{{Korshunov}}},
  \bibinfo{journal}{Europhys. Lett.} \textbf{\bibinfo{volume}{9}},
  \bibinfo{pages}{107} (\bibinfo{year}{1989}).

\bibitem[{\citenamefont{Roushan et~al.}(2017)\citenamefont{Roushan, Neill,
  Megrant, Chen, Babbush, Barends et~al.}}]{roushan2017}
\bibinfo{author}{\bibfnamefont{P.}~\bibnamefont{Roushan}},
  \bibinfo{author}{\bibfnamefont{C.}~\bibnamefont{Neill}},
  \bibinfo{author}{\bibfnamefont{A.}~\bibnamefont{Megrant}},
  \bibinfo{author}{\bibfnamefont{Y.}~\bibnamefont{Chen}},
  \bibinfo{author}{\bibfnamefont{R.}~\bibnamefont{Babbush}},
  \bibinfo{author}{\bibfnamefont{R.}~\bibnamefont{Barends}},
  \bibnamefont{et~al.}, \bibinfo{journal}{Nat. Phys.}
  \textbf{\bibinfo{volume}{13}}, \bibinfo{pages}{146} (\bibinfo{year}{2017}).

\bibitem[{\citenamefont{{Le Hur} et~al.}(2016)\citenamefont{{Le Hur}, Henriet,
  Petrescu, Plekhanov, Roux, and {Schir\'o}}}]{lehur2015}
\bibinfo{author}{\bibfnamefont{K.}~\bibnamefont{{Le Hur}}},
  \bibinfo{author}{\bibfnamefont{L.}~\bibnamefont{Henriet}},
  \bibinfo{author}{\bibfnamefont{A.}~\bibnamefont{Petrescu}},
  \bibinfo{author}{\bibfnamefont{K.}~\bibnamefont{Plekhanov}},
  \bibinfo{author}{\bibfnamefont{G.}~\bibnamefont{Roux}}, \bibnamefont{and}
  \bibinfo{author}{\bibfnamefont{M.}~\bibnamefont{{Schir\'o}}},
  \bibinfo{journal}{C. R. Phys.} \textbf{\bibinfo{volume}{17}},
  \bibinfo{pages}{808} (\bibinfo{year}{2016}), \eprint{arXiv:1505.00167}.

\bibitem[{\citenamefont{Romero et~al.}(2017)\citenamefont{Romero, Solano, and
  Lamata}}]{romero2017}
\bibinfo{author}{\bibfnamefont{G.}~\bibnamefont{Romero}},
  \bibinfo{author}{\bibfnamefont{E.}~\bibnamefont{Solano}}, \bibnamefont{and}
  \bibinfo{author}{\bibfnamefont{L.}~\bibnamefont{Lamata}}, in
  \emph{\bibinfo{booktitle}{Quantum Simulations with Photons and Polaritons:
  Merging Quantum Optics with Condensed Matter Physics}}, edited by
  \bibinfo{editor}{\bibfnamefont{D.}~\bibnamefont{Angelakis}}
  (\bibinfo{publisher}{Springer}, \bibinfo{address}{Heidelberg},
  \bibinfo{year}{2017}), Quantum Science and Technology Series,
  chap.~\bibinfo{chapter}{7}, p. \bibinfo{pages}{153}.

\bibitem[{\citenamefont{{Dhar} et~al.}(2012)\citenamefont{{Dhar}, {Maji},
  {Mishra}, {Pai}, {Mukerjee}, and {Paramekanti}}}]{dhar2012}
\bibinfo{author}{\bibfnamefont{A.}~\bibnamefont{{Dhar}}},
  \bibinfo{author}{\bibfnamefont{M.}~\bibnamefont{{Maji}}},
  \bibinfo{author}{\bibfnamefont{T.}~\bibnamefont{{Mishra}}},
  \bibinfo{author}{\bibfnamefont{R.~V.} \bibnamefont{{Pai}}},
  \bibinfo{author}{\bibfnamefont{S.}~\bibnamefont{{Mukerjee}}},
  \bibnamefont{and}
  \bibinfo{author}{\bibfnamefont{A.}~\bibnamefont{{Paramekanti}}},
  \bibinfo{journal}{Phys. Rev. A} \textbf{\bibinfo{volume}{85}},
  \bibinfo{pages}{041602} (\bibinfo{year}{2012}).

\bibitem[{\citenamefont{{Dhar} et~al.}(2013)\citenamefont{{Dhar}, {Mishra},
  {Maji}, {Pai}, {Mukerjee}, and {Paramekanti}}}]{dhar2013}
\bibinfo{author}{\bibfnamefont{A.}~\bibnamefont{{Dhar}}},
  \bibinfo{author}{\bibfnamefont{T.}~\bibnamefont{{Mishra}}},
  \bibinfo{author}{\bibfnamefont{M.}~\bibnamefont{{Maji}}},
  \bibinfo{author}{\bibfnamefont{R.~V.} \bibnamefont{{Pai}}},
  \bibinfo{author}{\bibfnamefont{S.}~\bibnamefont{{Mukerjee}}},
  \bibnamefont{and}
  \bibinfo{author}{\bibfnamefont{A.}~\bibnamefont{{Paramekanti}}},
  \bibinfo{journal}{Phys. Rev. B} \textbf{\bibinfo{volume}{87}},
  \bibinfo{pages}{174501} (\bibinfo{year}{2013}).

\bibitem[{\citenamefont{{Petrescu} and {Le Hur}}(2013)}]{petrescu2013}
\bibinfo{author}{\bibfnamefont{A.}~\bibnamefont{{Petrescu}}} \bibnamefont{and}
  \bibinfo{author}{\bibfnamefont{K.}~\bibnamefont{{Le Hur}}},
  \bibinfo{journal}{Phys. Rev. Lett.} \textbf{\bibinfo{volume}{111}},
  \bibinfo{pages}{150601} (\bibinfo{year}{2013}).

\bibitem[{\citenamefont{Po et~al.}(2014)\citenamefont{Po, Chen, and
  Zhou}}]{po2014}
\bibinfo{author}{\bibfnamefont{H.~C.} \bibnamefont{Po}},
  \bibinfo{author}{\bibfnamefont{W.}~\bibnamefont{Chen}}, \bibnamefont{and}
  \bibinfo{author}{\bibfnamefont{Q.}~\bibnamefont{Zhou}},
  \bibinfo{journal}{Phys. Rev. A} \textbf{\bibinfo{volume}{90}},
  \bibinfo{pages}{011602} (\bibinfo{year}{2014}).

\bibitem[{\citenamefont{Xu et~al.}(2014)\citenamefont{Xu, Cole, and
  Zhang}}]{xu2014}
\bibinfo{author}{\bibfnamefont{Z.}~\bibnamefont{Xu}},
  \bibinfo{author}{\bibfnamefont{W.}~\bibnamefont{Cole}}, \bibnamefont{and}
  \bibinfo{author}{\bibfnamefont{S.}~\bibnamefont{Zhang}},
  \bibinfo{journal}{Phys. Rev. A} \textbf{\bibinfo{volume}{89}},
  \bibinfo{pages}{051604(R)} (\bibinfo{year}{2014}), \eprint{arXiv:1403.3491}.

\bibitem[{\citenamefont{Zhao et~al.}(2014)\citenamefont{Zhao, Hu, Chang, Zheng,
  Zhang, and Wang}}]{zhao2014}
\bibinfo{author}{\bibfnamefont{J.}~\bibnamefont{Zhao}},
  \bibinfo{author}{\bibfnamefont{S.}~\bibnamefont{Hu}},
  \bibinfo{author}{\bibfnamefont{J.}~\bibnamefont{Chang}},
  \bibinfo{author}{\bibfnamefont{F.}~\bibnamefont{Zheng}},
  \bibinfo{author}{\bibfnamefont{P.}~\bibnamefont{Zhang}}, \bibnamefont{and}
  \bibinfo{author}{\bibfnamefont{X.}~\bibnamefont{Wang}},
  \bibinfo{journal}{Phys. Rev. B} \textbf{\bibinfo{volume}{90}},
  \bibinfo{pages}{085117} (\bibinfo{year}{2014}).

\bibitem[{\citenamefont{Wei and Mueller}(2014)}]{wei2014}
\bibinfo{author}{\bibfnamefont{R.}~\bibnamefont{Wei}} \bibnamefont{and}
  \bibinfo{author}{\bibfnamefont{E.~J.} \bibnamefont{Mueller}},
  \bibinfo{journal}{Phys. Rev. A} \textbf{\bibinfo{volume}{89}},
  \bibinfo{pages}{063617} (\bibinfo{year}{2014}).

\bibitem[{\citenamefont{H\"ugel and Paredes}(2014)}]{huegel2014}
\bibinfo{author}{\bibfnamefont{D.}~\bibnamefont{H\"ugel}} \bibnamefont{and}
  \bibinfo{author}{\bibfnamefont{B.}~\bibnamefont{Paredes}},
  \bibinfo{journal}{Phys. Rev. A} \textbf{\bibinfo{volume}{89}},
  \bibinfo{pages}{023619} (\bibinfo{year}{2014}).

\bibitem[{\citenamefont{Piraud et~al.}(2014)\citenamefont{Piraud, Cai,
  McCulloch, and Schollw{\"o}ck}}]{piraud2014}
\bibinfo{author}{\bibfnamefont{M.}~\bibnamefont{Piraud}},
  \bibinfo{author}{\bibfnamefont{Z.}~\bibnamefont{Cai}},
  \bibinfo{author}{\bibfnamefont{I.~P.} \bibnamefont{McCulloch}},
  \bibnamefont{and}
  \bibinfo{author}{\bibfnamefont{U.}~\bibnamefont{Schollw{\"o}ck}},
  \bibinfo{journal}{Phys. Rev. A} \textbf{\bibinfo{volume}{89}},
  \bibinfo{pages}{063618} (\bibinfo{year}{2014}).

\bibitem[{\citenamefont{Piraud et~al.}(2015)\citenamefont{Piraud,
  Heidrich-Meisner, McCulloch, Greschner, Vekua, and
  Schollw\"ock}}]{piraud2014b}
\bibinfo{author}{\bibfnamefont{M.}~\bibnamefont{Piraud}},
  \bibinfo{author}{\bibfnamefont{F.}~\bibnamefont{Heidrich-Meisner}},
  \bibinfo{author}{\bibfnamefont{I.~P.} \bibnamefont{McCulloch}},
  \bibinfo{author}{\bibfnamefont{S.}~\bibnamefont{Greschner}},
  \bibinfo{author}{\bibfnamefont{T.}~\bibnamefont{Vekua}}, \bibnamefont{and}
  \bibinfo{author}{\bibfnamefont{U.}~\bibnamefont{Schollw\"ock}},
  \bibinfo{journal}{Phys. Rev. B} \textbf{\bibinfo{volume}{91}},
  \bibinfo{pages}{140406} (\bibinfo{year}{2015}).

\bibitem[{\citenamefont{Kele\c{s} and Oktel}(2015)}]{keles2015}
\bibinfo{author}{\bibfnamefont{A.}~\bibnamefont{Kele\c{s}}} \bibnamefont{and}
  \bibinfo{author}{\bibfnamefont{M.~O.} \bibnamefont{Oktel}},
  \bibinfo{journal}{Phys. Rev. A} \textbf{\bibinfo{volume}{91}},
  \bibinfo{pages}{013629} (\bibinfo{year}{2015}).

\bibitem[{\citenamefont{Greschner et~al.}(2015)\citenamefont{Greschner, Piraud,
  Heidrich-Meisner, McCulloch, Schollw{\"o}ck, and Vekua}}]{greschner2015}
\bibinfo{author}{\bibfnamefont{S.}~\bibnamefont{Greschner}},
  \bibinfo{author}{\bibfnamefont{M.}~\bibnamefont{Piraud}},
  \bibinfo{author}{\bibfnamefont{F.}~\bibnamefont{Heidrich-Meisner}},
  \bibinfo{author}{\bibfnamefont{I.}~\bibnamefont{McCulloch}},
  \bibinfo{author}{\bibfnamefont{U.}~\bibnamefont{Schollw{\"o}ck}},
  \bibnamefont{and} \bibinfo{author}{\bibfnamefont{T.}~\bibnamefont{Vekua}},
  \bibinfo{journal}{Phys. Rev. Lett.} \textbf{\bibinfo{volume}{115}},
  \bibinfo{pages}{190402} (\bibinfo{year}{2015}).

\bibitem[{\citenamefont{Greschner et~al.}(2016)\citenamefont{Greschner, Piraud,
  Heidrich-Meisner, McCulloch, Schollw\"ock, and Vekua}}]{greschner2016}
\bibinfo{author}{\bibfnamefont{S.}~\bibnamefont{Greschner}},
  \bibinfo{author}{\bibfnamefont{M.}~\bibnamefont{Piraud}},
  \bibinfo{author}{\bibfnamefont{F.}~\bibnamefont{Heidrich-Meisner}},
  \bibinfo{author}{\bibfnamefont{I.~P.} \bibnamefont{McCulloch}},
  \bibinfo{author}{\bibfnamefont{U.}~\bibnamefont{Schollw\"ock}},
  \bibnamefont{and} \bibinfo{author}{\bibfnamefont{T.}~\bibnamefont{Vekua}},
  \bibinfo{journal}{Phys. Rev. A} \textbf{\bibinfo{volume}{94}},
  \bibinfo{pages}{063628} (\bibinfo{year}{2016}).

\bibitem[{\citenamefont{Petrescu and Le~Hur}(2015)}]{petrescu2015}
\bibinfo{author}{\bibfnamefont{A.}~\bibnamefont{Petrescu}} \bibnamefont{and}
  \bibinfo{author}{\bibfnamefont{K.}~\bibnamefont{Le~Hur}},
  \bibinfo{journal}{Phys. Rev. B} \textbf{\bibinfo{volume}{91}},
  \bibinfo{pages}{054520} (\bibinfo{year}{2015}).

\bibitem[{\citenamefont{Barbiero et~al.}(2016)\citenamefont{Barbiero, Abad, and
  Recati}}]{barbiero2014}
\bibinfo{author}{\bibfnamefont{L.}~\bibnamefont{Barbiero}},
  \bibinfo{author}{\bibfnamefont{M.}~\bibnamefont{Abad}}, \bibnamefont{and}
  \bibinfo{author}{\bibfnamefont{A.}~\bibnamefont{Recati}},
  \bibinfo{journal}{Phys. Rev. A} \textbf{\bibinfo{volume}{93}},
  \bibinfo{pages}{033645} (\bibinfo{year}{2016}), \eprint{arXiv:1403.4185}.

\bibitem[{\citenamefont{Peotta et~al.}(2014)\citenamefont{Peotta, Mazza,
  Vicari, Polini, Fazio, and Rossini}}]{peotta2014}
\bibinfo{author}{\bibfnamefont{S.}~\bibnamefont{Peotta}},
  \bibinfo{author}{\bibfnamefont{L.}~\bibnamefont{Mazza}},
  \bibinfo{author}{\bibfnamefont{E.}~\bibnamefont{Vicari}},
  \bibinfo{author}{\bibfnamefont{M.}~\bibnamefont{Polini}},
  \bibinfo{author}{\bibfnamefont{R.}~\bibnamefont{Fazio}}, \bibnamefont{and}
  \bibinfo{author}{\bibfnamefont{D.}~\bibnamefont{Rossini}},
  \bibinfo{journal}{J. Stat. Mech.: Theory Exp.}
  \textbf{\bibinfo{volume}{2014}}, \bibinfo{pages}{P09005}
  (\bibinfo{year}{2014}).

\bibitem[{\citenamefont{Di~Dio et~al.}(2015)\citenamefont{Di~Dio, De~Palo,
  Orignac, Citro, and Chiofalo}}]{our_2015}
\bibinfo{author}{\bibfnamefont{M.}~\bibnamefont{Di~Dio}},
  \bibinfo{author}{\bibfnamefont{S.}~\bibnamefont{De~Palo}},
  \bibinfo{author}{\bibfnamefont{E.}~\bibnamefont{Orignac}},
  \bibinfo{author}{\bibfnamefont{R.}~\bibnamefont{Citro}}, \bibnamefont{and}
  \bibinfo{author}{\bibfnamefont{M.-L.} \bibnamefont{Chiofalo}},
  \bibinfo{journal}{Phys. Rev. B} \textbf{\bibinfo{volume}{92}},
  \bibinfo{pages}{060506} (\bibinfo{year}{2015}).

\bibitem[{\citenamefont{Orignac et~al.}(2016)\citenamefont{Orignac, Citro,
  Di~Dio, De~Palo, and Chiofalo}}]{our_2016}
\bibinfo{author}{\bibfnamefont{E.}~\bibnamefont{Orignac}},
  \bibinfo{author}{\bibfnamefont{R.}~\bibnamefont{Citro}},
  \bibinfo{author}{\bibfnamefont{M.}~\bibnamefont{Di~Dio}},
  \bibinfo{author}{\bibfnamefont{S.}~\bibnamefont{De~Palo}}, \bibnamefont{and}
  \bibinfo{author}{\bibfnamefont{M.~L.} \bibnamefont{Chiofalo}},
  \bibinfo{journal}{New J. Phys.} \textbf{\bibinfo{volume}{18}},
  \bibinfo{pages}{055017} (\bibinfo{year}{2016}).

\bibitem[{\citenamefont{Petrescu et~al.}(2016)\citenamefont{Petrescu, Piraud,
  Roux, McCulloch, and Hur}}]{petrescu2016}
\bibinfo{author}{\bibfnamefont{A.}~\bibnamefont{Petrescu}},
  \bibinfo{author}{\bibfnamefont{M.}~\bibnamefont{Piraud}},
  \bibinfo{author}{\bibfnamefont{G.}~\bibnamefont{Roux}},
  \bibinfo{author}{\bibfnamefont{I.}~\bibnamefont{McCulloch}},
  \bibnamefont{and} \bibinfo{author}{\bibfnamefont{K.~L.} \bibnamefont{Hur}},
  \emph{\bibinfo{title}{Precursor of laughlin state of hard core bosons on a
  two leg ladder}}, \bibinfo{howpublished}{arXiv:1612.05134}
  (\bibinfo{year}{2016}).

\bibitem[{\citenamefont{Barbarino et~al.}(2016)\citenamefont{Barbarino, Taddia,
  Rossini, Mazza, and Fazio}}]{barbarino2016}
\bibinfo{author}{\bibfnamefont{S.}~\bibnamefont{Barbarino}},
  \bibinfo{author}{\bibfnamefont{L.}~\bibnamefont{Taddia}},
  \bibinfo{author}{\bibfnamefont{D.}~\bibnamefont{Rossini}},
  \bibinfo{author}{\bibfnamefont{L.}~\bibnamefont{Mazza}}, \bibnamefont{and}
  \bibinfo{author}{\bibfnamefont{R.}~\bibnamefont{Fazio}},
  \bibinfo{journal}{New J. Phys.} \textbf{\bibinfo{volume}{18}},
  \bibinfo{pages}{035010} (\bibinfo{year}{2016}).

\bibitem[{\citenamefont{Strinati et~al.}(2017)\citenamefont{Strinati, Cornfeld,
  Rossini, Barbarino, Dalmonte, Fazio, Sela, and Mazza}}]{strinati2017}
\bibinfo{author}{\bibfnamefont{M.~C.} \bibnamefont{Strinati}},
  \bibinfo{author}{\bibfnamefont{E.}~\bibnamefont{Cornfeld}},
  \bibinfo{author}{\bibfnamefont{D.}~\bibnamefont{Rossini}},
  \bibinfo{author}{\bibfnamefont{S.}~\bibnamefont{Barbarino}},
  \bibinfo{author}{\bibfnamefont{M.}~\bibnamefont{Dalmonte}},
  \bibinfo{author}{\bibfnamefont{R.}~\bibnamefont{Fazio}},
  \bibinfo{author}{\bibfnamefont{E.}~\bibnamefont{Sela}}, \bibnamefont{and}
  \bibinfo{author}{\bibfnamefont{L.}~\bibnamefont{Mazza}},
  \bibinfo{journal}{Phys. Rev. X} \textbf{\bibinfo{volume}{7}},
  \bibinfo{pages}{021033} (\bibinfo{year}{2017}).

\bibitem[{\citenamefont{Tokuno and Georges}(2014)}]{tokuno2014}
\bibinfo{author}{\bibfnamefont{A.}~\bibnamefont{Tokuno}} \bibnamefont{and}
  \bibinfo{author}{\bibfnamefont{A.}~\bibnamefont{Georges}},
  \bibinfo{journal}{New J. Phys.} \textbf{\bibinfo{volume}{16}},
  \bibinfo{pages}{073005} (\bibinfo{year}{2014}).

\bibitem[{\citenamefont{Uchino and Tokuno}(2015)}]{uchino2015}
\bibinfo{author}{\bibfnamefont{S.}~\bibnamefont{Uchino}} \bibnamefont{and}
  \bibinfo{author}{\bibfnamefont{A.}~\bibnamefont{Tokuno}},
  \bibinfo{journal}{Phys. Rev. A} \textbf{\bibinfo{volume}{92}},
  \bibinfo{pages}{013625} (\bibinfo{year}{2015}).

\bibitem[{\citenamefont{Bilitewski and Cooper}(2016)}]{bilitewski2016}
\bibinfo{author}{\bibfnamefont{T.}~\bibnamefont{Bilitewski}} \bibnamefont{and}
  \bibinfo{author}{\bibfnamefont{N.~R.} \bibnamefont{Cooper}},
  \bibinfo{journal}{Phys. Rev. A} \textbf{\bibinfo{volume}{94}},
  \bibinfo{pages}{023630} (\bibinfo{year}{2016}).

\bibitem[{\citenamefont{Greschner and Vekua}(2017)}]{greschner2017}
\bibinfo{author}{\bibfnamefont{S.}~\bibnamefont{Greschner}} \bibnamefont{and}
  \bibinfo{author}{\bibfnamefont{T.}~\bibnamefont{Vekua}},
  \emph{\bibinfo{title}{Vortex-hole duality: a unified picture of weak and
  strong-coupling regimes of bosonic ladders with flux}},
  \bibinfo{howpublished}{arXiv:1704.06517} (\bibinfo{year}{2017}).

\bibitem[{\citenamefont{Guo and Poletti}(2017)}]{guo2017}
\bibinfo{author}{\bibfnamefont{C.}~\bibnamefont{Guo}} \bibnamefont{and}
  \bibinfo{author}{\bibfnamefont{D.}~\bibnamefont{Poletti}},
  \emph{\bibinfo{title}{Dissipatively driven strongly interacting bosons in a
  gauge field}}, \bibinfo{howpublished}{arXiv preprint arXiv:1705.07633}
  (\bibinfo{year}{2017}).

\bibitem[{\citenamefont{Richaud and Penna}(2017)}]{richaud2017}
\bibinfo{author}{\bibfnamefont{A.}~\bibnamefont{Richaud}} \bibnamefont{and}
  \bibinfo{author}{\bibfnamefont{V.}~\bibnamefont{Penna}},
  \emph{\bibinfo{title}{Quantum dynamics of bosons in a two-ring ladder:
  dynamical algebra, vortex-like excitations and currents}},
  \bibinfo{howpublished}{arXiv preprint arXiv:1705.02115}
  (\bibinfo{year}{2017}).

\bibitem[{\citenamefont{{Romen} and {L{\"a}uchli}}(2017)}]{romen2017}
\bibinfo{author}{\bibfnamefont{C.}~\bibnamefont{{Romen}}} \bibnamefont{and}
  \bibinfo{author}{\bibfnamefont{A.~M.} \bibnamefont{{L{\"a}uchli}}},
  \emph{\bibinfo{title}{{Chiral Mott insulators in frustrated Bose-Hubbard
  models on ladders and two-dimensional lattices: a combined perturbative and
  density matrix renormalization group study}}} (\bibinfo{year}{2017}),
  \eprint{arXiv:1711.01909}.

\bibitem[{\citenamefont{Laughlin}(1983)}]{laughlin_FQHE}
\bibinfo{author}{\bibfnamefont{R.~B.} \bibnamefont{Laughlin}},
  \bibinfo{journal}{Phys. Rev. Lett.} \textbf{\bibinfo{volume}{50}},
  \bibinfo{pages}{1395} (\bibinfo{year}{1983}).

\bibitem[{\citenamefont{Natu}(2015)}]{natu2015}
\bibinfo{author}{\bibfnamefont{S.~S.} \bibnamefont{Natu}},
  \emph{\bibinfo{title}{Bosons with long range interactions on two-leg ladders
  in artificial magnetic fields}} (\bibinfo{year}{2015}),
  \bibinfo{note}{arXiv:1506.04346}.

\bibitem[{\citenamefont{Orignac et~al.}(2017)\citenamefont{Orignac, Citro,
  {Di~Dio}, and {De~Palo}}}]{orignac2017}
\bibinfo{author}{\bibfnamefont{E.}~\bibnamefont{Orignac}},
  \bibinfo{author}{\bibfnamefont{R.}~\bibnamefont{Citro}},
  \bibinfo{author}{\bibfnamefont{M.}~\bibnamefont{{Di~Dio}}}, \bibnamefont{and}
  \bibinfo{author}{\bibfnamefont{S.}~\bibnamefont{{De~Palo}}},
  \bibinfo{journal}{Phys. Rev. B} \textbf{\bibinfo{volume}{96}},
  \bibinfo{pages}{014518} (\bibinfo{year}{2017}),
  \bibinfo{note}{arXiv:1703.07742}.

\bibitem[{\citenamefont{{Bohr} et~al.}(1982)\citenamefont{{Bohr}, {Pokrovski{\v
  i}}, and {Talapov}}}]{bohr1982}
\bibinfo{author}{\bibfnamefont{T.}~\bibnamefont{{Bohr}}},
  \bibinfo{author}{\bibfnamefont{V.~L.} \bibnamefont{{Pokrovski{\v i}}}},
  \bibnamefont{and} \bibinfo{author}{\bibfnamefont{A.~L.}
  \bibnamefont{{Talapov}}}, \bibinfo{journal}{JETP Lett.}
  \textbf{\bibinfo{volume}{35}}, \bibinfo{pages}{203} (\bibinfo{year}{1982}).

\bibitem[{\citenamefont{Bohr}(1982)}]{bohr1982b}
\bibinfo{author}{\bibfnamefont{T.}~\bibnamefont{Bohr}}, \bibinfo{journal}{Phys.
  Rev. B} \textbf{\bibinfo{volume}{25}}, \bibinfo{pages}{6981}
  (\bibinfo{year}{1982}), \bibinfo{note}{[Phys. Rev. B \textbf{26}, 5257(E)
  (1982)]}.

\bibitem[{\citenamefont{Haldane et~al.}(1983)\citenamefont{Haldane, Bak, and
  Bohr}}]{haldane83_cic}
\bibinfo{author}{\bibfnamefont{F.~D.~M.} \bibnamefont{Haldane}},
  \bibinfo{author}{\bibfnamefont{P.}~\bibnamefont{Bak}}, \bibnamefont{and}
  \bibinfo{author}{\bibfnamefont{T.}~\bibnamefont{Bohr}},
  \bibinfo{journal}{Phys. Rev. B} \textbf{\bibinfo{volume}{28}},
  \bibinfo{pages}{2743} (\bibinfo{year}{1983}).

\bibitem[{\citenamefont{Schulz}(1983)}]{schulz83_cic_vortices}
\bibinfo{author}{\bibfnamefont{H.-J.} \bibnamefont{Schulz}},
  \bibinfo{journal}{Phys. Rev. B}
  \textbf{\bibinfo{volume}{28}}(\bibinfo{number}{5}), \bibinfo{pages}{2746}
  (\bibinfo{year}{1983}).

\bibitem[{\citenamefont{Horowitz et~al.}(1983)\citenamefont{Horowitz, Bohr,
  Kosterlitz, and Schulz}}]{horowitz_renormalization_incommensurable}
\bibinfo{author}{\bibfnamefont{B.}~\bibnamefont{Horowitz}},
  \bibinfo{author}{\bibfnamefont{T.}~\bibnamefont{Bohr}},
  \bibinfo{author}{\bibfnamefont{J.}~\bibnamefont{Kosterlitz}},
  \bibnamefont{and} \bibinfo{author}{\bibfnamefont{H.~J.}
  \bibnamefont{Schulz}}, \bibinfo{journal}{Phys. Rev. B}
  \textbf{\bibinfo{volume}{28}}, \bibinfo{pages}{6596} (\bibinfo{year}{1983}).

\bibitem[{\citenamefont{Stephenson}(1970{\natexlab{a}})}]{stephenson1970a}
\bibinfo{author}{\bibfnamefont{J.}~\bibnamefont{Stephenson}},
  \bibinfo{journal}{Can. J. Phys.} \textbf{\bibinfo{volume}{48}},
  \bibinfo{pages}{1724} (\bibinfo{year}{1970}{\natexlab{a}}).

\bibitem[{\citenamefont{Stephenson}(1970{\natexlab{b}})}]{stephenson1970b}
\bibinfo{author}{\bibfnamefont{J.}~\bibnamefont{Stephenson}},
  \bibinfo{journal}{Phys. Rev. B} \textbf{\bibinfo{volume}{1}},
  \bibinfo{pages}{4405} (\bibinfo{year}{1970}{\natexlab{b}}).

\bibitem[{\citenamefont{Berezinskii}(1971)}]{berezinskii_2dxy}
\bibinfo{author}{\bibfnamefont{V.~L.} \bibnamefont{Berezinskii}},
  \bibinfo{journal}{Sov. Phys. JETP} \textbf{\bibinfo{volume}{32}},
  \bibinfo{pages}{493} (\bibinfo{year}{1971}).

\bibitem[{\citenamefont{Kosterlitz and Thouless}(1973)}]{kosterlitz_thouless}
\bibinfo{author}{\bibfnamefont{J.~M.} \bibnamefont{Kosterlitz}}
  \bibnamefont{and} \bibinfo{author}{\bibfnamefont{D.~J.}
  \bibnamefont{Thouless}}, \bibinfo{journal}{J. Phys. C}
  \textbf{\bibinfo{volume}{6}}, \bibinfo{pages}{1181} (\bibinfo{year}{1973}).

\bibitem[{\citenamefont{Cazalilla and Ho}(2003)}]{cazalilla03_mixture}
\bibinfo{author}{\bibfnamefont{M.~A.} \bibnamefont{Cazalilla}}
  \bibnamefont{and} \bibinfo{author}{\bibfnamefont{A.~F.} \bibnamefont{Ho}},
  \bibinfo{journal}{Phys. Rev. Lett.} \textbf{\bibinfo{volume}{91}},
  \bibinfo{pages}{150403} (\bibinfo{year}{2003}).

\bibitem[{\citenamefont{Mathey et~al.}(2000)\citenamefont{Mathey, Danshita, and
  Clark}}]{mathey08_supersolid}
\bibinfo{author}{\bibfnamefont{L.}~\bibnamefont{Mathey}},
  \bibinfo{author}{\bibfnamefont{I.}~\bibnamefont{Danshita}}, \bibnamefont{and}
  \bibinfo{author}{\bibfnamefont{C.~W.} \bibnamefont{Clark}},
  \bibinfo{journal}{Phys. Rev. A} \textbf{\bibinfo{volume}{79}},
  \bibinfo{pages}{011602(R)} (\bibinfo{year}{2000}).

\bibitem[{\citenamefont{Hu et~al.}(2009)\citenamefont{Hu, Mathey, Danshita,
  Tiesinga, Williams, and Clark}}]{hu_pra_2009}
\bibinfo{author}{\bibfnamefont{A.}~\bibnamefont{Hu}},
  \bibinfo{author}{\bibfnamefont{L.}~\bibnamefont{Mathey}},
  \bibinfo{author}{\bibfnamefont{I.}~\bibnamefont{Danshita}},
  \bibinfo{author}{\bibfnamefont{E.}~\bibnamefont{Tiesinga}},
  \bibinfo{author}{\bibfnamefont{C.~J.} \bibnamefont{Williams}},
  \bibnamefont{and} \bibinfo{author}{\bibfnamefont{C.~W.} \bibnamefont{Clark}},
  \bibinfo{journal}{Phys. Rev. A} \textbf{\bibinfo{volume}{80}},
  \bibinfo{pages}{023619} (\bibinfo{year}{2009}).

\bibitem[{\citenamefont{Haldane}(1981)}]{haldane_bosons}
\bibinfo{author}{\bibfnamefont{F.~D.~M.} \bibnamefont{Haldane}},
  \bibinfo{journal}{Phys. Rev. Lett.} \textbf{\bibinfo{volume}{47}},
  \bibinfo{pages}{1840} (\bibinfo{year}{1981}).

\bibitem[{\citenamefont{Lukyanov and Terras}(2003)}]{lukyanov_xxz_asymptotics}
\bibinfo{author}{\bibfnamefont{S.}~\bibnamefont{Lukyanov}} \bibnamefont{and}
  \bibinfo{author}{\bibfnamefont{V.}~\bibnamefont{Terras}},
  \bibinfo{journal}{Nucl. Phys. B} \textbf{\bibinfo{volume}{654}},
  \bibinfo{pages}{323} (\bibinfo{year}{2003}), \bibinfo{note}{hep-th/0206093}.

\bibitem[{\citenamefont{{Ovchinnikov}}(2004)}]{ovchinnikov2004}
\bibinfo{author}{\bibfnamefont{A.~A.} \bibnamefont{{Ovchinnikov}}},
  \bibinfo{journal}{J. Phys.: Condens. Matter} \textbf{\bibinfo{volume}{16}},
  \bibinfo{pages}{3147} (\bibinfo{year}{2004}), \eprint{arXiv:math-ph/0311050}.

\bibitem[{\citenamefont{Shashi et~al.}(2012)\citenamefont{Shashi, Panfil, Caux,
  and Imambekov}}]{shashi2012}
\bibinfo{author}{\bibfnamefont{A.}~\bibnamefont{Shashi}},
  \bibinfo{author}{\bibfnamefont{M.}~\bibnamefont{Panfil}},
  \bibinfo{author}{\bibfnamefont{J.-S.} \bibnamefont{Caux}}, \bibnamefont{and}
  \bibinfo{author}{\bibfnamefont{A.}~\bibnamefont{Imambekov}},
  \bibinfo{journal}{Phys. Rev. B} \textbf{\bibinfo{volume}{85}},
  \bibinfo{pages}{155136} (\bibinfo{year}{2012}).

\bibitem[{\citenamefont{Hikihara and
  Furusaki}(2003)}]{hikihara03_amplitude_xxz}
\bibinfo{author}{\bibfnamefont{T.}~\bibnamefont{Hikihara}} \bibnamefont{and}
  \bibinfo{author}{\bibfnamefont{A.}~\bibnamefont{Furusaki}},
  \emph{\bibinfo{title}{Correlation amplitudes for the spin-1/2 xxz chain in a
  magnetic field}} (\bibinfo{year}{2003}), \bibinfo{note}{cond-mat/0310391}.

\bibitem[{\citenamefont{Bouillot et~al.}(2011)\citenamefont{Bouillot, Kollath,
  Läuchli, Zvonarev, Thielemann, Rüegg, Orignac, Citro, Klanjsek, Berthier
  et~al.}}]{bouillot2010}
\bibinfo{author}{\bibfnamefont{P.}~\bibnamefont{Bouillot}},
  \bibinfo{author}{\bibfnamefont{C.}~\bibnamefont{Kollath}},
  \bibinfo{author}{\bibfnamefont{A.~M.} \bibnamefont{Läuchli}},
  \bibinfo{author}{\bibfnamefont{M.}~\bibnamefont{Zvonarev}},
  \bibinfo{author}{\bibfnamefont{B.}~\bibnamefont{Thielemann}},
  \bibinfo{author}{\bibfnamefont{C.}~\bibnamefont{Rüegg}},
  \bibinfo{author}{\bibfnamefont{E.}~\bibnamefont{Orignac}},
  \bibinfo{author}{\bibfnamefont{R.}~\bibnamefont{Citro}},
  \bibinfo{author}{\bibfnamefont{M.}~\bibnamefont{Klanjsek}},
  \bibinfo{author}{\bibfnamefont{C.}~\bibnamefont{Berthier}},
  \bibnamefont{et~al.}, \bibinfo{journal}{Phys. Rev. B}
  \textbf{\bibinfo{volume}{83}}, \bibinfo{pages}{054407}
  (\bibinfo{year}{2011}), \bibinfo{note}{arXiv:1009.0840}.

\bibitem[{\citenamefont{Giamarchi}(2004)}]{giamarchi_book_1d}
\bibinfo{author}{\bibfnamefont{T.}~\bibnamefont{Giamarchi}},
  \emph{\bibinfo{title}{Quantum Physics in One Dimension}}
  (\bibinfo{publisher}{Oxford University Press}, \bibinfo{address}{Oxford},
  \bibinfo{year}{2004}).

\bibitem[{\citenamefont{Orignac and Giamarchi}(1998)}]{orignac98_vortices}
\bibinfo{author}{\bibfnamefont{E.}~\bibnamefont{Orignac}} \bibnamefont{and}
  \bibinfo{author}{\bibfnamefont{T.}~\bibnamefont{Giamarchi}},
  \bibinfo{journal}{Phys. Rev. B} \textbf{\bibinfo{volume}{57}},
  \bibinfo{pages}{11713} (\bibinfo{year}{1998}), \eprint{cond-mat/9801048}.

\bibitem[{\citenamefont{Mathey}(2007)}]{mathey07_mixture}
\bibinfo{author}{\bibfnamefont{L.}~\bibnamefont{Mathey}},
  \bibinfo{journal}{Phys. Rev. B} \textbf{\bibinfo{volume}{75}},
  \bibinfo{pages}{144510} (\bibinfo{year}{2007}).

\bibitem[{\citenamefont{Mathey and Wang}(2007)}]{mathey07_bose_fermi}
\bibinfo{author}{\bibfnamefont{L.}~\bibnamefont{Mathey}} \bibnamefont{and}
  \bibinfo{author}{\bibfnamefont{D.-W.} \bibnamefont{Wang}},
  \bibinfo{journal}{Phys. Rev. A} \textbf{\bibinfo{volume}{75}},
  \bibinfo{pages}{013602} (\bibinfo{year}{2007}).

\bibitem[{\citenamefont{Mathey et~al.}(2009)\citenamefont{Mathey, Danshita, and
  Clark}}]{mathey_pra_2009}
\bibinfo{author}{\bibfnamefont{L.}~\bibnamefont{Mathey}},
  \bibinfo{author}{\bibfnamefont{I.}~\bibnamefont{Danshita}}, \bibnamefont{and}
  \bibinfo{author}{\bibfnamefont{C.~W.} \bibnamefont{Clark}},
  \bibinfo{journal}{Phys. Rev. A} \textbf{\bibinfo{volume}{79}},
  \bibinfo{pages}{011602} (\bibinfo{year}{2009}).

\bibitem[{\citenamefont{Jos{\'e} et~al.}(1977)\citenamefont{Jos{\'e}, Kadanoff,
  Kirkpatrick, and Nelson}}]{jose_planar_2d}
\bibinfo{author}{\bibfnamefont{J.~V.} \bibnamefont{Jos{\'e}}},
  \bibinfo{author}{\bibfnamefont{L.~P.} \bibnamefont{Kadanoff}},
  \bibinfo{author}{\bibfnamefont{S.}~\bibnamefont{Kirkpatrick}},
  \bibnamefont{and} \bibinfo{author}{\bibfnamefont{D.~R.}
  \bibnamefont{Nelson}}, \bibinfo{journal}{Phys. Rev. B}
  \textbf{\bibinfo{volume}{16}}, \bibinfo{pages}{1217} (\bibinfo{year}{1977}).

\bibitem[{\citenamefont{Lecheminant et~al.}(2002)\citenamefont{Lecheminant,
  Gogolin, and Nersesyan}}]{lecheminant2002sdsg}
\bibinfo{author}{\bibfnamefont{P.}~\bibnamefont{Lecheminant}},
  \bibinfo{author}{\bibfnamefont{A.~O.} \bibnamefont{Gogolin}},
  \bibnamefont{and} \bibinfo{author}{\bibfnamefont{A.~A.}
  \bibnamefont{Nersesyan}}, \bibinfo{journal}{Nucl. Phys. B}
  \textbf{\bibinfo{volume}{639}}, \bibinfo{pages}{502} (\bibinfo{year}{2002}).

\bibitem[{\citenamefont{Japaridze and
  Nersesyan}(1978)}]{japaridze_cic_transition}
\bibinfo{author}{\bibfnamefont{G.~I.} \bibnamefont{Japaridze}}
  \bibnamefont{and} \bibinfo{author}{\bibfnamefont{A.~A.}
  \bibnamefont{Nersesyan}}, \bibinfo{journal}{JETP Lett.}
  \textbf{\bibinfo{volume}{27}}, \bibinfo{pages}{334} (\bibinfo{year}{1978}).

\bibitem[{\citenamefont{Pokrovsky and Talapov}(1979)}]{pokrovsky_talapov_prl}
\bibinfo{author}{\bibfnamefont{V.~L.} \bibnamefont{Pokrovsky}}
  \bibnamefont{and} \bibinfo{author}{\bibfnamefont{A.~L.}
  \bibnamefont{Talapov}}, \bibinfo{journal}{Phys. Rev. Lett.}
  \textbf{\bibinfo{volume}{42}}, \bibinfo{pages}{65} (\bibinfo{year}{1979}).

\bibitem[{\citenamefont{Schulz}(1980)}]{schulz_cic2d}
\bibinfo{author}{\bibfnamefont{H.~J.} \bibnamefont{Schulz}},
  \bibinfo{journal}{Phys. Rev. B} \textbf{\bibinfo{volume}{22}},
  \bibinfo{pages}{5274} (\bibinfo{year}{1980}).

\bibitem[{\citenamefont{Chitra and Giamarchi}(1997)}]{chitra_spinchains_field}
\bibinfo{author}{\bibfnamefont{R.}~\bibnamefont{Chitra}} \bibnamefont{and}
  \bibinfo{author}{\bibfnamefont{T.}~\bibnamefont{Giamarchi}},
  \bibinfo{journal}{Phys. Rev. B} \textbf{\bibinfo{volume}{55}},
  \bibinfo{pages}{5816} (\bibinfo{year}{1997}).

\bibitem[{\citenamefont{Orignac and
  Giamarchi}(2001{\natexlab{b}})}]{orignac_vortex_ladder}
\bibinfo{author}{\bibfnamefont{E.}~\bibnamefont{Orignac}} \bibnamefont{and}
  \bibinfo{author}{\bibfnamefont{T.}~\bibnamefont{Giamarchi}},
  \bibinfo{journal}{Phys. Rev. B} \textbf{\bibinfo{volume}{64}},
  \bibinfo{pages}{144515} (\bibinfo{year}{2001}{\natexlab{b}}).

\bibitem[{\citenamefont{Tsvelik}(1990)}]{tsvelik_field}
\bibinfo{author}{\bibfnamefont{A.~M.} \bibnamefont{Tsvelik}},
  \bibinfo{journal}{Phys. Rev. B} \textbf{\bibinfo{volume}{42}},
  \bibinfo{pages}{10499} (\bibinfo{year}{1990}).

\bibitem[{\citenamefont{Wang}(2003)}]{wang2003field}
\bibinfo{author}{\bibfnamefont{Y.-J.} \bibnamefont{Wang}},
  \emph{\bibinfo{title}{Field-induced ising criticality and incommensurability
  in anisotropic spin-1 chains}}, \bibinfo{howpublished}{arXiv preprint
  cond-mat/0306365} (\bibinfo{year}{2003}).

\bibitem[{\citenamefont{Essler and Affleck}(2004)}]{essler04_spin1_field}
\bibinfo{author}{\bibfnamefont{F.~H.~L.} \bibnamefont{Essler}}
  \bibnamefont{and} \bibinfo{author}{\bibfnamefont{I.}~\bibnamefont{Affleck}},
  \bibinfo{journal}{J. Stat. Mech.: Theory Exp.} p. \bibinfo{pages}{P12006}
  (\bibinfo{year}{2004}).

\bibitem[{\citenamefont{Shelton et~al.}(1996)\citenamefont{Shelton, Nersesyan,
  and Tsvelik}}]{shelton_spin_ladders}
\bibinfo{author}{\bibfnamefont{D.~G.} \bibnamefont{Shelton}},
  \bibinfo{author}{\bibfnamefont{A.~A.} \bibnamefont{Nersesyan}},
  \bibnamefont{and} \bibinfo{author}{\bibfnamefont{A.~M.}
  \bibnamefont{Tsvelik}}, \bibinfo{journal}{Phys. Rev. B}
  \textbf{\bibinfo{volume}{53}}, \bibinfo{pages}{8521} (\bibinfo{year}{1996}).

\bibitem[{\citenamefont{Nersesyan and Tsvelik}(1997)}]{nersesyan_biquad}
\bibinfo{author}{\bibfnamefont{A.}~\bibnamefont{Nersesyan}} \bibnamefont{and}
  \bibinfo{author}{\bibfnamefont{A.~M.} \bibnamefont{Tsvelik}},
  \bibinfo{journal}{Phys. Rev. Lett.} \textbf{\bibinfo{volume}{78}},
  \bibinfo{pages}{3939} (\bibinfo{year}{1997}), \bibinfo{note}{ibid. ,
  \textbf{79}, E 1171}.

\bibitem[{\citenamefont{Citro and Orignac}(2002)}]{citro02_dm_ladders}
\bibinfo{author}{\bibfnamefont{R.}~\bibnamefont{Citro}} \bibnamefont{and}
  \bibinfo{author}{\bibfnamefont{E.}~\bibnamefont{Orignac}},
  \bibinfo{journal}{Phys. Rev. B} \textbf{\bibinfo{volume}{65}},
  \bibinfo{pages}{134413} (\bibinfo{year}{2002}), \eprint{cond-mat/0106020}.

\bibitem[{\citenamefont{McCoy}(1995)}]{mccoy_revue_qft}
\bibinfo{author}{\bibfnamefont{B.~M.} \bibnamefont{McCoy}}, in
  \emph{\bibinfo{booktitle}{Statistical Mechanics and Field Theory}}, edited by
  \bibinfo{editor}{\bibfnamefont{V.}~\bibnamefont{Bazhanov}} \bibnamefont{and}
  \bibinfo{editor}{\bibfnamefont{C.}~\bibnamefont{Burden}}
  (\bibinfo{publisher}{World Scientific}, \bibinfo{address}{Singapore},
  \bibinfo{year}{1995}), p.~\bibinfo{pages}{26},
  \bibinfo{note}{hep-th/9403084}.

\bibitem[{\citenamefont{Campostrini et~al.}(2014)\citenamefont{Campostrini,
  Pelissetto, and Vicari}}]{campostrini2014}
\bibinfo{author}{\bibfnamefont{M.}~\bibnamefont{Campostrini}},
  \bibinfo{author}{\bibfnamefont{A.}~\bibnamefont{Pelissetto}},
  \bibnamefont{and} \bibinfo{author}{\bibfnamefont{E.}~\bibnamefont{Vicari}},
  \bibinfo{journal}{Phys. Rev. B} \textbf{\bibinfo{volume}{89}},
  \bibinfo{pages}{094516} (\bibinfo{year}{2014}).

\bibitem[{\citenamefont{Zuber and Itzykson}(1977)}]{zuber_77}
\bibinfo{author}{\bibfnamefont{J.~B.} \bibnamefont{Zuber}} \bibnamefont{and}
  \bibinfo{author}{\bibfnamefont{C.}~\bibnamefont{Itzykson}},
  \bibinfo{journal}{Phys. Rev. D} \textbf{\bibinfo{volume}{15}},
  \bibinfo{pages}{2875} (\bibinfo{year}{1977}).

\bibitem[{\citenamefont{Schroer and Truong}(1978)}]{schroer_ising}
\bibinfo{author}{\bibfnamefont{B.}~\bibnamefont{Schroer}} \bibnamefont{and}
  \bibinfo{author}{\bibfnamefont{T.~T.} \bibnamefont{Truong}},
  \bibinfo{journal}{Nucl. Phys. B} \textbf{\bibinfo{volume}{144}},
  \bibinfo{pages}{80} (\bibinfo{year}{1978}).

\bibitem[{\citenamefont{Boyanovsky}(1989)}]{boyanovsky_ising}
\bibinfo{author}{\bibfnamefont{D.}~\bibnamefont{Boyanovsky}},
  \bibinfo{journal}{Phys. Rev. B} \textbf{\bibinfo{volume}{39}},
  \bibinfo{pages}{6744} (\bibinfo{year}{1989}).

\bibitem[{\citenamefont{Nersesyan}(2001)}]{nersesyan2001_ising}
\bibinfo{author}{\bibfnamefont{A.~A.} \bibnamefont{Nersesyan}}, in
  \emph{\bibinfo{booktitle}{New theoretical approaches to strongly correlated
  systems}}, edited by \bibinfo{editor}{\bibfnamefont{A.~M.}
  \bibnamefont{Tsvelik}} (\bibinfo{publisher}{Kluwer Academic Publishers},
  \bibinfo{address}{Dordrecht, Netherlands}, \bibinfo{year}{2001}),
  vol.~\bibinfo{volume}{23} of \emph{\bibinfo{series}{NATO science series.
  Series II, Mathematics, physics, and chemistry}}, chap.~\bibinfo{chapter}{4},
  p.~\bibinfo{pages}{89}.

\bibitem[{\citenamefont{M.Bender and Orszag}(1978)}]{bender78_book}
\bibinfo{author}{\bibfnamefont{C.}~\bibnamefont{M.Bender}} \bibnamefont{and}
  \bibinfo{author}{\bibfnamefont{S.~A.} \bibnamefont{Orszag}},
  \emph{\bibinfo{title}{Advanced mathematical methods for scientists and
  engineers}} (\bibinfo{publisher}{McGraw-Hill}, \bibinfo{address}{NY},
  \bibinfo{year}{1978}).

\bibitem[{\citenamefont{Sondhi et~al.}(1997)\citenamefont{Sondhi, Girvin,
  Carini, and Shahar}}]{sondhi_qcp}
\bibinfo{author}{\bibfnamefont{L.}~\bibnamefont{Sondhi}},
  \bibinfo{author}{\bibfnamefont{S.~M.} \bibnamefont{Girvin}},
  \bibinfo{author}{\bibfnamefont{J.~P.} \bibnamefont{Carini}},
  \bibnamefont{and} \bibinfo{author}{\bibfnamefont{D.}~\bibnamefont{Shahar}},
  \bibinfo{journal}{Rev. Mod. Phys.} \textbf{\bibinfo{volume}{69}},
  \bibinfo{pages}{315} (\bibinfo{year}{1997}).

\bibitem[{\citenamefont{Sachdev}(2000)}]{sachdev_book}
\bibinfo{author}{\bibfnamefont{S.}~\bibnamefont{Sachdev}},
  \emph{\bibinfo{title}{Quantum Phase Transitions}}
  (\bibinfo{publisher}{Cambridge University Press},
  \bibinfo{address}{Cambridge, UK}, \bibinfo{year}{2000}).

\bibitem[{\citenamefont{Landau and Lifshitz}(1986)}]{landau-statmech-english}
\bibinfo{author}{\bibfnamefont{L.~D.} \bibnamefont{Landau}} \bibnamefont{and}
  \bibinfo{author}{\bibfnamefont{I.~M.} \bibnamefont{Lifshitz}},
  \emph{\bibinfo{title}{Statistical Physics. {\rm 3rd edition}}}
  (\bibinfo{publisher}{Pergamon Press}, \bibinfo{address}{Oxford},
  \bibinfo{year}{1986}).

\bibitem[{\citenamefont{Sachdev}(1996)}]{sachdev_ising}
\bibinfo{author}{\bibfnamefont{S.}~\bibnamefont{Sachdev}},
  \bibinfo{journal}{Nucl. Phys. B} \textbf{\bibinfo{volume}{464}},
  \bibinfo{pages}{576} (\bibinfo{year}{1996}).

\bibitem[{\citenamefont{Sachdev and Young}(1997)}]{sachdev1997}
\bibinfo{author}{\bibfnamefont{S.}~\bibnamefont{Sachdev}} \bibnamefont{and}
  \bibinfo{author}{\bibfnamefont{A.~P.} \bibnamefont{Young}},
  \bibinfo{journal}{Phys. Rev. Lett.} \textbf{\bibinfo{volume}{78}},
  \bibinfo{pages}{2220} (\bibinfo{year}{1997}),
  \urlprefix\url{http://link.aps.org/doi/10.1103/PhysRevLett.78.2220}.

\bibitem[{\citenamefont{Giamarchi and Schulz}(1988)}]{giamarchi_spin_flop}
\bibinfo{author}{\bibfnamefont{T.}~\bibnamefont{Giamarchi}} \bibnamefont{and}
  \bibinfo{author}{\bibfnamefont{H.~J.} \bibnamefont{Schulz}},
  \bibinfo{journal}{J. Phys. (Paris)} \textbf{\bibinfo{volume}{49}},
  \bibinfo{pages}{819} (\bibinfo{year}{1988}).

\bibitem[{\citenamefont{Nersesyan et~al.}(1993)\citenamefont{Nersesyan, Luther,
  and Kusmartsev}}]{nersesyan_2ch}
\bibinfo{author}{\bibfnamefont{A.}~\bibnamefont{Nersesyan}},
  \bibinfo{author}{\bibfnamefont{A.}~\bibnamefont{Luther}}, \bibnamefont{and}
  \bibinfo{author}{\bibfnamefont{F.}~\bibnamefont{Kusmartsev}},
  \bibinfo{journal}{Phys. Lett. A} \textbf{\bibinfo{volume}{176}},
  \bibinfo{pages}{363} (\bibinfo{year}{1993}).

\bibitem[{\citenamefont{{Di Dio} et~al.}(2015)\citenamefont{{Di Dio}, Citro,
  {De Palo}, Orignac, and Chiofalo}}]{didio2015a}
\bibinfo{author}{\bibfnamefont{M.}~\bibnamefont{{Di Dio}}},
  \bibinfo{author}{\bibfnamefont{R.}~\bibnamefont{Citro}},
  \bibinfo{author}{\bibfnamefont{S.}~\bibnamefont{{De Palo}}},
  \bibinfo{author}{\bibfnamefont{E.}~\bibnamefont{Orignac}}, \bibnamefont{and}
  \bibinfo{author}{\bibfnamefont{M.-L.} \bibnamefont{Chiofalo}},
  \bibinfo{journal}{Eur. Phys. J. Spec. Top.} \textbf{\bibinfo{volume}{224}},
  \bibinfo{pages}{525} (\bibinfo{year}{2015}).

\bibitem[{\citenamefont{Dzyaloshinskii}(1958)}]{dzyaloshinskii_interaction}
\bibinfo{author}{\bibfnamefont{I.}~\bibnamefont{Dzyaloshinskii}},
  \bibinfo{journal}{J. Phys. Chem. Solids} \textbf{\bibinfo{volume}{4}},
  \bibinfo{pages}{241} (\bibinfo{year}{1958}).

\bibitem[{\citenamefont{Moriya}(1960)}]{moryia_asym_int}
\bibinfo{author}{\bibfnamefont{T.}~\bibnamefont{Moriya}},
  \bibinfo{journal}{Phys. Rev.} \textbf{\bibinfo{volume}{120}},
  \bibinfo{pages}{91} (\bibinfo{year}{1960}).

\bibitem[{\citenamefont{Holstein and Primakoff}(1940)}]{holstein40_operators}
\bibinfo{author}{\bibfnamefont{T.}~\bibnamefont{Holstein}} \bibnamefont{and}
  \bibinfo{author}{\bibfnamefont{H.}~\bibnamefont{Primakoff}},
  \bibinfo{journal}{Phys. Rev.} \textbf{\bibinfo{volume}{58}},
  \bibinfo{pages}{1098} (\bibinfo{year}{1940}).

\bibitem[{\citenamefont{Lieb and Wu}(1968)}]{lieb_hubbard_exact}
\bibinfo{author}{\bibfnamefont{E.~H.} \bibnamefont{Lieb}} \bibnamefont{and}
  \bibinfo{author}{\bibfnamefont{F.~Y.} \bibnamefont{Wu}},
  \bibinfo{journal}{Phys. Rev. Lett.} \textbf{\bibinfo{volume}{20}},
  \bibinfo{pages}{1445} (\bibinfo{year}{1968}).

\bibitem[{\citenamefont{Frahm and Korepin}(1990)}]{frahm_confinv}
\bibinfo{author}{\bibfnamefont{H.}~\bibnamefont{Frahm}} \bibnamefont{and}
  \bibinfo{author}{\bibfnamefont{V.~E.} \bibnamefont{Korepin}},
  \bibinfo{journal}{Phys. Rev. B} \textbf{\bibinfo{volume}{42}},
  \bibinfo{pages}{10553} (\bibinfo{year}{1990}).

\bibitem[{\citenamefont{Frahm and Korepin}(1991)}]{frahm_confinv_field}
\bibinfo{author}{\bibfnamefont{H.}~\bibnamefont{Frahm}} \bibnamefont{and}
  \bibinfo{author}{\bibfnamefont{V.~E.} \bibnamefont{Korepin}},
  \bibinfo{journal}{Phys. Rev. B} \textbf{\bibinfo{volume}{43}},
  \bibinfo{pages}{5663} (\bibinfo{year}{1991}).

\bibitem[{\citenamefont{Andrei}(1993)}]{andrei_trieste93}
\bibinfo{author}{\bibfnamefont{N.}~\bibnamefont{Andrei}}, in
  \emph{\bibinfo{booktitle}{Low-Dimensional Quantum Field Theories For
  Condensed Matter Physicists}}, edited by
  \bibinfo{editor}{\bibfnamefont{S.}~\bibnamefont{Lundqvist}},
  \bibinfo{editor}{\bibfnamefont{G.}~\bibnamefont{Morandi}}, \bibnamefont{and}
  \bibinfo{editor}{\bibfnamefont{L.}~\bibnamefont{Yu}}
  (\bibinfo{publisher}{World Scientific}, \bibinfo{address}{Singapore},
  \bibinfo{year}{1993}), \bibinfo{note}{and references therein}.

\bibitem[{\citenamefont{Jordan and Wigner}(1928)}]{jordan_transformation}
\bibinfo{author}{\bibfnamefont{P.}~\bibnamefont{Jordan}} \bibnamefont{and}
  \bibinfo{author}{\bibfnamefont{E.}~\bibnamefont{Wigner}},
  \bibinfo{journal}{Z. Phys.} \textbf{\bibinfo{volume}{47}},
  \bibinfo{pages}{631} (\bibinfo{year}{1928}).

\bibitem[{\citenamefont{Zvyagin}(2012)}]{zvyagin2012}
\bibinfo{author}{\bibfnamefont{A.~A.} \bibnamefont{Zvyagin}},
  \bibinfo{journal}{Phys. Rev. B} \textbf{\bibinfo{volume}{86}},
  \bibinfo{pages}{085126} (\bibinfo{year}{2012}).

\bibitem[{\citenamefont{Beccaria et~al.}(2006)\citenamefont{Beccaria,
  Campostrini, and Feo}}]{beccaria2006}
\bibinfo{author}{\bibfnamefont{M.}~\bibnamefont{Beccaria}},
  \bibinfo{author}{\bibfnamefont{M.}~\bibnamefont{Campostrini}},
  \bibnamefont{and} \bibinfo{author}{\bibfnamefont{A.}~\bibnamefont{Feo}},
  \bibinfo{journal}{Phys. Rev. B} \textbf{\bibinfo{volume}{73}},
  \bibinfo{pages}{052402} (\bibinfo{year}{2006}).

\bibitem[{\citenamefont{Golinelli et~al.}(1998)\citenamefont{Golinelli,
  Jolicoeur, and Sorensen}}]{golinelli_incommensurate}
\bibinfo{author}{\bibfnamefont{O.}~\bibnamefont{Golinelli}},
  \bibinfo{author}{\bibfnamefont{T.}~\bibnamefont{Jolicoeur}},
  \bibnamefont{and} \bibinfo{author}{\bibfnamefont{E.}~\bibnamefont{Sorensen}},
  \emph{\bibinfo{title}{Incommensurability in the magnetic excitations of the
  bilinear-biquadratic spin-1 chain}} (\bibinfo{year}{1998}),
  \bibinfo{note}{cond-mat/9812296}.

\bibitem[{\citenamefont{Schollw{\"o}ck
  et~al.}(1996)\citenamefont{Schollw{\"o}ck, Jolicoeur, and
  Garel}}]{schollwoeck1996}
\bibinfo{author}{\bibfnamefont{U.}~\bibnamefont{Schollw{\"o}ck}},
  \bibinfo{author}{\bibfnamefont{T.}~\bibnamefont{Jolicoeur}},
  \bibnamefont{and} \bibinfo{author}{\bibfnamefont{T.}~\bibnamefont{Garel}},
  \bibinfo{journal}{Phys. Rev. B} \textbf{\bibinfo{volume}{53}},
  \bibinfo{pages}{3304} (\bibinfo{year}{1996}).

\bibitem[{\citenamefont{Bursill et~al.}(1995)\citenamefont{Bursill, Gehring,
  Farnell, Parkinson, Xiang, and Zeng}}]{bursill1995}
\bibinfo{author}{\bibfnamefont{R.}~\bibnamefont{Bursill}},
  \bibinfo{author}{\bibfnamefont{G.}~\bibnamefont{Gehring}},
  \bibinfo{author}{\bibfnamefont{D.}~\bibnamefont{Farnell}},
  \bibinfo{author}{\bibfnamefont{J.}~\bibnamefont{Parkinson}},
  \bibinfo{author}{\bibfnamefont{T.}~\bibnamefont{Xiang}}, \bibnamefont{and}
  \bibinfo{author}{\bibfnamefont{C.}~\bibnamefont{Zeng}}, \bibinfo{journal}{J.
  Phys.: Condens. Matter} \textbf{\bibinfo{volume}{7}}, \bibinfo{pages}{8605}
  (\bibinfo{year}{1995}).

\bibitem[{\citenamefont{Deschner and S{\o}rensen}(2013)}]{deschner2013}
\bibinfo{author}{\bibfnamefont{A.}~\bibnamefont{Deschner}} \bibnamefont{and}
  \bibinfo{author}{\bibfnamefont{E.~S.} \bibnamefont{S{\o}rensen}},
  \bibinfo{journal}{Phys. Rev. B} \textbf{\bibinfo{volume}{87}},
  \bibinfo{pages}{094415} (\bibinfo{year}{2013}).

\bibitem[{\citenamefont{Pixley et~al.}(2014)\citenamefont{Pixley, Shashi, and
  Nevidomskyy}}]{pixley2014}
\bibinfo{author}{\bibfnamefont{J.}~\bibnamefont{Pixley}},
  \bibinfo{author}{\bibfnamefont{A.}~\bibnamefont{Shashi}}, \bibnamefont{and}
  \bibinfo{author}{\bibfnamefont{A.~H.} \bibnamefont{Nevidomskyy}},
  \bibinfo{journal}{Physical Review B} \textbf{\bibinfo{volume}{90}},
  \bibinfo{pages}{214426} (\bibinfo{year}{2014}).

\bibitem[{\citenamefont{Chepiga et~al.}(2016)\citenamefont{Chepiga, Affleck,
  and Mila}}]{chepiga2016}
\bibinfo{author}{\bibfnamefont{N.}~\bibnamefont{Chepiga}},
  \bibinfo{author}{\bibfnamefont{I.}~\bibnamefont{Affleck}}, \bibnamefont{and}
  \bibinfo{author}{\bibfnamefont{F.}~\bibnamefont{Mila}},
  \bibinfo{journal}{Phys. Rev. B} \textbf{\bibinfo{volume}{94}},
  \bibinfo{pages}{205112} (\bibinfo{year}{2016}).

\bibitem[{\citenamefont{Witten}(1984)}]{witten_wz}
\bibinfo{author}{\bibfnamefont{E.}~\bibnamefont{Witten}},
  \bibinfo{journal}{Commun . Math. Phys.} \textbf{\bibinfo{volume}{92}},
  \bibinfo{pages}{455} (\bibinfo{year}{1984}).

\end{thebibliography}

\end{document}